\documentclass[twocolumn, trackchanges]{aastex631}

\usepackage{txfonts}
\usepackage{xspace}
\usepackage{mathrsfs}
\usepackage{amssymb}
\usepackage{gensymb}
\newcommand{\nnhp}{$\rm N_2H^+$\xspace}
\newcommand{\hhdp}{$\rm H_2D^+$\xspace}
\newcommand{\nndp}{$\rm N_2D^+$\xspace}

\newcommand{\dcop}{$\rm DCO^+$\xspace}

\newcommand{\kms}{$\rm km \, s^{-1}$\xspace}
\newcommand{\ncol}{$N_\mathrm{col}$\xspace}

\newcommand{\tex}{$T_\mathrm{ex}$\xspace}
\newcommand{\vlsr}{$V_\mathrm{lsr}$\xspace}
\newcommand{\mcore}{$M_\mathrm{core}$\xspace}
\newcommand{\sigmav}{$\sigma_\mathrm{V}$\xspace}

\newcommand{\ohhdp}{$\rm \text{o-} H_2D^+$\xspace}

\newcommand{\olineh}{$\rm \text{o-} H_2D^+(1_{1,0} - 1_{1,1})$\xspace}

\shorttitle{An ALMA view of the high-mass clump AG14}
\shortauthors{Redaelli et al.}
\graphicspath{{./}{figures/}}

\begin{document}

\title{The core population and kinematics of a massive clump at early stages: an ALMA view}

\correspondingauthor{Elena Redaelli}
\email{eredaelli@mpe.mpg.de}

\author[0000-0002-0528-8125]{Elena Redaelli}
\affiliation{Centre for Astrochemical Studies, Max-Planck-Institut f\"ur extraterrestrische Physik, Gie{\ss}enbachstra{\ss}e 1, 85749 Garching bei M\"unchen, Germany}

\author{Stefano Bovino}
\affiliation{Departamento de Astronom\'ia, Facultad Ciencias F\'isicas y Matem\'aticas, Universidad de Concepci\'on, Av. Esteban Iturra s/n Barrio
Universitario, Casilla 160, Concepci\'on, Chile}
\affiliation{INAF --- Istituto di Radioastronomia --- Italian node of the ALMA Regional Centre (It-ARC), Via Gobetti 101, 40129 Bologna, Italy}

\author{Patricio Sanhueza}
\affiliation{National Astronomical Observatory of Japan, National Institutes of Natural Sciences, 2-21-1 Osawa, Mitaka, Tokyo 181-8588, Japan}
\affiliation{Department of Astronomical Science, The Graduate University for Advanced Studies, SOKENDAI, 2-21-1 Osawa, Mitaka, Tokyo 181-8588, Japan f}

\author{Kaho Morii}
\affiliation{National Astronomical Observatory of Japan, National Institutes of Natural Sciences, 2-21-1 Osawa, Mitaka, Tokyo 181-8588, Japan}
\affiliation{Department of Astronomy, Graduate School of Science, The University of Tokyo, 7-3-1 Hongo, Bunkyo-ku, Tokyo 113-0033, Japan}

\author{Giovanni Sabatini}
\affiliation{INAF --- Istituto di Radioastronomia --- Italian node of the ALMA Regional Centre (It-ARC), Via Gobetti 101, 40129 Bologna, Italy}

\author{Paola Caselli}
\affiliation{Centre for Astrochemical Studies, Max-Planck-Institut f\"ur extraterrestrische Physik, Gie{\ss}enbachstra{\ss}e 1, 85749 Garching bei M\"unchen, Germany}

\author{Andrea Giannetti}
\affiliation{ INAF - Istituto di Radioastronomia, Via P. Gobetti 101, I-40129 Bologna, Italy}
\author{Shanghuo Li} 
\affiliation{Korea Astronomy and Space Science Institute, 776 Daedeokdae-ro, Yuseong-gu, Daejeon 34055, Republic Of Korea}


\begin{abstract}

 High-mass star formation theories make distinct predictions on the properties of the prestellar seeds of high-mass stars. Observations of the early stages of high-mass star formation can provide crucial constraints, but they are challenging and scarce. We investigate the properties of the prestellar core population embedded in the high-mass clump  AGAL014.492-00.139, and we study the kinematics at the clump and the clump-to-core scales. We have analysed an extensive dataset acquired with the ALMA interferometer. Applying a dendrogram analysis to the Band 7 \ohhdp data, we identified 22 cores. We have fitted their average spectra in local-thermodinamic-equilibrium conditions, and we analysed their continuum emission at $0.8 \, \rm mm$. The cores have transonic to mildly supersonic turbulence levels and appear mostly low-mass, with $M_\mathrm{core}< 30 \, \rm M_\odot$. Furthermore, we have analysed Band 3 observations of the \nnhp (1-0) transition, which traces the large scale gas kinematics. Using a friend-of-friend algorithm, we identify four main velocity coherent structures, all of which are associated with prestellar and protostellar cores. One of them presents a filament-like structure, {and our observations could be consistent with mass accretion towards one of the protostars. In this case, we} estimate a mass accretion rate of $ \dot{M}_\mathrm{acc}\approx 2 \times 10^{-4} \rm \, M_\odot \, yr^{-1}$. Our results support a clump-fed accretion scenario in the targeted source. The cores in  prestellar stage are essentially low-mass, and they appear subvirial and gravitationally bound, unless further support is available for instance due to magnetic fields. 
\end{abstract}

\keywords{Interstellar line emission(844) --- Star forming regions(1565) --- Astrochemistry(75) --- Interferometry(808) ---Massive stars(732) --- Star formation(1569) }


\section{Introduction\label{Intro}}
High-mass stars dominate the energetic of the interstellar medium (ISM), mainly due to feedback during their whole life cycle. Despite their importance, however, their formation process is significantly less known that the low-mass counterpart. From the theoretical point of view, two main families of models  have been developed. The core-accretion (or core-fed) model is a scaled-up version with respect to the low-mass process \citep{McKee03}. It predicts the existence of high mass prestellar cores (HMPCs, $M_\mathrm{core}=$ several tens of solar masses), which are virialised either due to turbulence or to the contribution of magnetic pressure, that collapse as a whole \citep{Tan13, Tan14}. In the clump-fed or competitive accretion scenarios, instead, early fragmentation in high-mass clumps leads to the formation of essentially low-mass cores, which keep accreting mass from the dense surrounding environment also during the initial protostellar stages \citep{Bonnel01, Bonnell06, Smith09}. In order to distinguish among the existing theories, observational constraints on the properties of the initial stages of high-mass star formation are needed, in particular in terms of core masses and properties of accretion. \par
These observations are however challenging, since high-mass stars are intrinsically rarer and on average more distant than low-mass ones. The birthplace of high-mass stars is to be found in the heavily obscured environments of infrared dark clouds (IRDCs, \citealt{Rathborne06}). In particular, IRDCs that are dark at $24$ and $70 \, \rm \mu m$ are supposed to host the earliest evolutionary stages of high-mass star formation \citep{Tan13, Sanhueza13,Guzman15}. Several studies have hence targeted IRDCs with interferometric facilities, such as the Atacama Large Millimeter/submillimeter Array (ALMA, as done by \citealt{Zhang15, Ohashi16, Contreras18, Svoboda19, Sanhueza19, Morii21}), or the Submillimeter Array \citep[SMA, see for instance][]{Sanhueza17, Li19, Pillai19}. {Multiple works} unveiled that the lack of emission at mid-infrared wavelengths as seen with single-dish facilities (e.g. the {\em Spitzer Space Telescope}) does not guarantee a complete lack of star formation activity, due to the high extinction that characterises high-mass star-forming regions \citep[see e.g.][]{Tan16, Pillai19, Li20, Morii21,Tafoya21}. \par
In this context, the ALMA Survey of $70 \, \mu$m dark High-mass clumps in Early Stages survey \citep[ASHES;][]{Sanhueza19} targeted twelve IRDCs with ALMA Band 6 observations. In the first paper of the series, the authors studied the clumps fragmentation using the continuum emission at  $1.3 \,$mm, identifying $\approx 300$ cores, none of which appears more massive than $30 \, \rm M_\odot$. Continuum emission together with spectral line observations have the potential to provide a more complete picture of star-forming regions, in particular in terms of evolutionary stage assessment. For instance, outflow tracers (e.g. CO, SiO), or so-called warm transitions, which have high upper level energies ($E_\mathrm{u}> 20-30 \, \rm K$), can be used to identify signs of protostellar activity, such as outflow emission or gas heating \citep[see e,g,][]{Sanhueza12, Li20}. \par
In the hunt for HMPCs, it is crucial to find a good and unambiguous tracer of the prestellar phases. Deuterated species appear promising to this aim. At low temperatures ($T < 20\, \rm K$) and high densities ($n \gtrsim 10^5 \, \rm cm^{-3})$ found in prestellar gas, most C- and O-bearing species are frozen out onto dust grains \citep{Caselli99,Bacmann02}. This contributes to increasing the abundance of \hhdp, the precursor of deuterated species in the gas phase, since this molecule is predominantly destroyed by reaction with CO \citep[e.g.][and references therein]{Ceccarelli14}. This results in a boost of deuteration, and deuterated molecules can therefore be good probes of cold and dense gas.
\par
\cite{Redaelli21} reported the first \ohhdp observations with ALMA in high-mass star-forming regions{ and showed that} this molecule is a good probe of prestellar conditions. 
{The \olineh line was detected towards two intermediate-mass clumps (AG351 and AG354)}, at a spatial resolution of $\approx 1500\, $AU. {The authors identified 16 cores in total, and estimated their masses from the continuum emission at $0.8\rm \,  mm$.} {At} $T_\mathrm{dust} = 10 \, $K, all cores are less massive than $10\, \rm M_\odot$, and the majority is subvirial, assuming negligible contribution to the stability from magnetic fields.
\par
Molecular lines yield information also on the gas kinematics, which is of great importance when trying to investigate the accretion processes in high-mass clumps. Among the different tracers used, two important ones are ammonia (see e.g. \citealt{ Lu18, Williams18, Sokolov19}) and \nnhp \citep{Henshaw14, Chen19}. {The kinematics of high-mass star-forming regions can be studied by means of algorithms dedicated to identify the hierarchy in their filamentary structures, as done for instance by \cite{Peretto14, Chen19, Henshaw19, Wang20}. Many of these works report the detection of velocity gradients usually interpreted as gas motions, linked to accretion flows towards cores or hubs \citep[see for instance][and references therein]{Hacar22}.}

\par
The $70\, \mu \rm m$-dark clump AGAL014.492-00.139 (hereafter AG14) has an estimated mass of $5200 \, \rm M_\odot$ and it is located at a distance of $3.9\, \rm kpc$ \citep{Sanhueza19}. It belongs to the ATLASGAL TOP100 sample \citep{Giannetti14,Konig17}, a statistically significant sample of high-mass clumps at different evolutionary stages in the inner Galaxy. AG14 was also included among the targets of the ASHES project: \cite{Sanhueza19} identified 37 cores in continuum, 25 of which are associated with warm line or outflow emission. This point was investigated further by \cite{Li20}, who used $\rm CO$ and $\rm SiO$ observations with ALMA, and found that six cores are associated with outflows. In particular, four present bipolar emission. Throughout this work, we will refer to these cores as protostellar (or protostars).  More recently, \cite{Sakai22} studied the emission of several deuterated molecules (\nndp, \dcop, and DCN) found in ALMA Band 6.
\par
In this work, we present an extensive ALMA dataset on AG14, from $90$ up to $370 \rm \, GHz$ in Sect. \ref{Observations}{, consisting of Band 3 data covering the \nnhp (1-0) line, Band 7 data of the \olineh line, and Band 6 data of the \nndp (3-2) transition (already published in \citealt{Sakai22}). These different lines are used to trace distinct parts of the clump. \hhdp is mainly destroyed by reactions with CO, and it is hence sensitive to temperature rising beyond the CO desorption temperature ($\approx 20 \, \rm K$). Furthermore, its \olineh transition has a critical density of $n_\mathrm{c} \approx 10^5 \, \rm cm^{-3}$ \citep{Hugo09}, hence this line is an ideal tracer of cold and dense gas at the core scales. \nnhp is also a well known high-density tracer. Its first rotational transition has a critical density of $6 \times 10^4 \rm \, cm^{-3}$
, and it presents an isolated hyperfine component well separated from the others also in cases of large linewidths ($\sigma_\mathrm{V} \lesssim 2-4 \,$\kms). This component is usually optically thin or only moderately optically thick \citep{Sanhueza12, Barnes18, Fontani21}. In the intracloud gas in high-mass clumps, the \nnhp transition is excited over large scales. For all these reasons, \nnhp represents an ideal probe of the clump and clump-to-core kinematics. Finally, \nndp is also a high-density tracer, but \cite{Giannetti19} studied the correlation between the \olineh and the \nndp (3-2) transitions in three clumps embedded in the G351.77-0.51 complex, using single-dish data from APEX. The main result of those authors was an anticorrelation between the abundances of the two molecular species. This was explained as an evolutionary effect: in the prestellar phase, as time evolves, the abundance of \ohhdp is expected to lower, mainly due the conversion to its doubly and triply deuterated forms (see also \citealt{Sabatini20}). \nndp instead forms later, and then its abundance keep increasing, since it can be formed also from $\rm D_2H^+$ and $\rm D_3 ^+$ \citep[see for instance the chemical model of][]{Sipila13,Sipila15}. These findings hinted to the possibility of using the abundance ratio between \nndp and \ohhdp as an evolutionary indicator, and we aim to investigate this point in AG14 with the available data.}
\par
{The paper is organised as follows. The observations are presented in Sect. \ref{Observations}.} In the analysis, we first investigate  the core population embedded in the clump, using the \olineh data (Sect. \ref{CorePop}). We then present the clump-to-core kinematic properties in Sect. \ref{kinematics}, based on the analysis of \nnhp (1-0) data. In Sect. \ref{h2dp_n2dp} we analyse the correlation between the \ohhdp and the \nndp emission in the identified cores, and Sect. \ref{Conclusion} contains a discussion and the concluding remarks of this work. 

\section{Observations\label{Observations}}
The observations {used in this work} are described in the following {subsections}, and the main technical details (e.g. angular resolution, sensitivity,...) are summarised in Table \ref{ObsData}. If the data have already been published, we refer to the corresponding publication.

\begin{deluxetable*}{ccccc}
\tablecaption{Observational parameters\label{ObsData}}
\tablehead{ \colhead{Observation}                   & \colhead{Beam size\tablenotemark{a}}            & \colhead{Spatial res.}         & \colhead{$rms$}                  & \colhead{Spectral res.}        
}
\startdata                             
 \multicolumn{5}{c}{Band 7}                                                              \\
Continuum $0.5\,\rm mm$                     &       $0''.66\times 0''.50$, $PA=-73.4$\degree     &  $2600 \rm AU \times 2000 AU$  &      $0.8 \, \rm mJy/beam$       &        -              \\
\olineh                         &    $0''.67\times 0''.50$, $PA=-73.4$\degree          &     $2600 \rm AU \times 2000  AU$   &      $100 \, \rm mK$ &   $0.20\,$\kms        \\
\hline
                               \multicolumn{5}{c}{Band 6\tablenotemark{b}}   \\
 Continuum $1.34\,\rm mm$                         &       $1''.29\times 0''.85$, $PA=72.5$\degree     &  $5000 \rm AU \times 3300 AU$  &      $0.17 \, \rm mJy/beam$       &        -              \\                                                                                        
\nndp (3-2)                      &    $1''.44\times 1''.00$, $PA=74.8$\degree          &     $5600 \rm AU \times 3900 AU$   &      $180 \, \rm mK$ &   $0.17\,$\kms        \\
                   \hline
                                \multicolumn{5}{c}{Band 3}                                                              \\
 \nnhp (1-0)                         &    $2''.86\times 1''.61$, $PA=74.7$\degree          &     $11200 \rm AU \times 6200  AU$   &      $110 \, \rm mK$ &   $0.20\,$\kms        \\                               
\enddata
\tablenotetext{a}{The beam size is shown as: major axis $\times$ minor axis, position angle ($PA$).}
\tablenotetext{b}{Data presented in \cite{Sanhueza19} and \cite{Sakai22}.}
\end{deluxetable*}

\subsection{Band 7 observations}
The Band 7 data were observed during Cycle 6 as part of the ALMA project 2018.1.00331.S (PI: Bovino) in three runs (November 2018 and March-April 2019). The observations, performed as {a} single-pointing, made use of both the Main Array (12m-array, 45 antennas) and the 7m-array (12 antennas), with baselines ranging from $7$ to $645 \, \rm m$. The quasars J1924-2914, J1911-2006, J1733-1304, and J1751+0939 were used as calibrators.  The spectral setup comprises four spectral windows (SPWs). The first one, dedicated to the observation of the \olineh transition, is centred at the frequency $\nu_\mathrm{rest} = 372421.3558\, \rm MHz$ \citep{Jusko17}, and has a resolution of $244 \, \rm kHz$ (corresponding to $0.20\,$\kms at  $372 \, \rm GHz$) and a total bandwidth of $500 \, \rm MHz$. The second SPW is dedicated to continuum, with a total bandwidth of $1.85 \, \rm GHz$ around the frequency of $371\, \rm GHz$. 
\par
At these frequencies, and with the used configuration, the maximum recoverable scale is $\theta_\mathrm{MRS} \approx 20''$, {the primary beams of the main array and of ACA are $17''$ and $30''$, respectively}, and the angular resolution is $\approx 0''.6$ (corresponding to $\approx 2300 \, \rm AU$ at the distance of $3.9\, \rm kpc$). The total observing time was $6.0\, \rm h$  ({7m-array}) and $2.5 \, \rm h$ (12m-array). During the observations, the precipitable water vapour was typically $0.4 \, \mathrm{mm}< PWV < 0.6\, \mathrm{mm}$. The average system temperature values are found in the range $300-400\, \rm K$ for the SPW containing the \olineh line. The data were calibrated by the standard pipeline (\textsc{casa}, version 5.4; \citealt{McMullin07}). From a first inspection of the dirty maps, the emission both in continuum and in line appear very extended in the whole Field-of-View (FoV). We therefore applied a modified weight of $2.4$ to the ACA observations, similarly to what was done in \cite{Redaelli21}{. After a few tests, this choice appeared the ideal compromise to maximise the recovery of the large-scale flux, without downgrading too much the final angular resolution.} \par 
We imaged the data using the \texttt{tclean} task of the software \textsc{casa} (version 5.6), in interactive mode. We used the natural weighting and the multiscale deconvolver algorithm \citep{Cornwell08} (scales: $0, 5, 15$). In order to avoid oversampling, both the continuum and the line images have been re-gridded in order to ensure 3 pixels per beam minor axis, in agreement with the Nyquist theorem. Table \ref{ObsData} summarises the achieved sensitivities and resolutions. The molecular line data have been converted into the brightness temperature $T_\mathrm{b}$ scale, using the gain {G computed as:} 
\begin{equation}
G = 1.222 \times 10^6 \frac{1}{\nu^2 \theta_\mathrm{min} \theta_\mathrm{maj}} = 26 \, \rm mK/ (mJy\, beam^{-1}) \;,
\label{gain_eq}
\end{equation}
{where $\nu$ is the frequency in GHz and $\theta_\mathrm{min/maj}$ are the beam sizes along the minor and major axes, respectively, expressed in arcsec.}
\par
\begin{figure}[h]
\centering
\includegraphics[width=.5\textwidth]{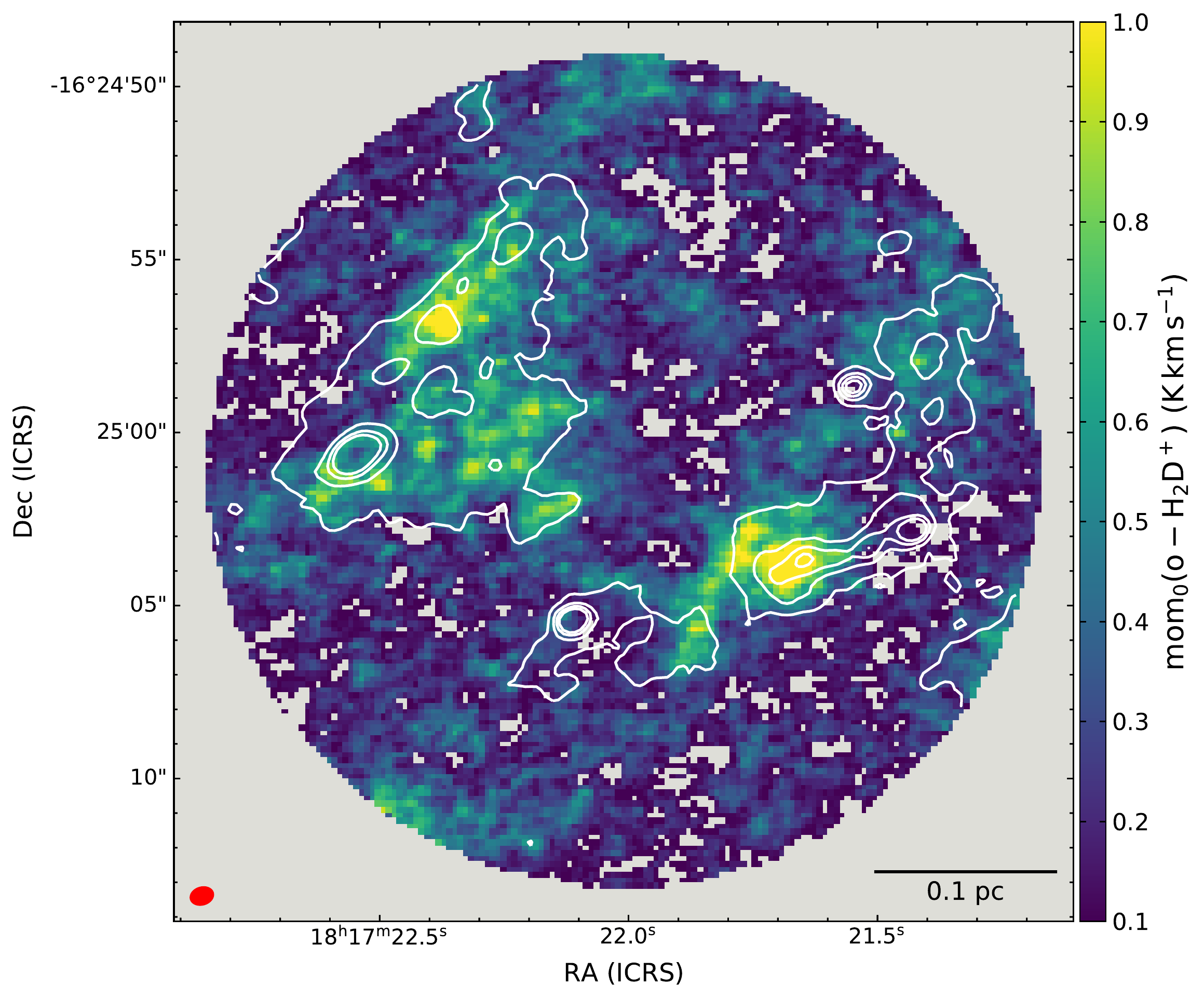}
\caption{{The colorscale shows the integrated intensity of the \olineh line, where pixels below the $3\sigma$ level have been masked. The mean uncertainty on the map is $1\sigma =60\, \rm mK\,$\kms. The contours show the continuum emission at levels from $2$ to $11\sigma$, in steps of $3\sigma$}. Both continuum and line data are shown prior to primary beam correction. The beam size and scalebar are indicated in the bottom left and right corners, respectively. \label{Band7_Data}}
\end{figure}

{Figure \ref{Band7_Data} shows the integrated intensity map of the \ohhdp line, computed in the velocity range $36-43\,$\kms, masking channels with a signal lower than $1\sigma$. The contours show the distribution of the continuum emission at $0.8\,\rm mm$.} Similarly to what has been noticed in \cite{Redaelli21} {in two different sources}, the morphology of the continuum and of the line emission are in general different. Several bright peaks identified in dust thermal emission lack a counterpart in \ohhdp emission above the $3\sigma$ level. 

\subsection{Band 6 observations}
The Band 6 data of the continuum emission and of the \nndp (3-2) line at $\nu_\mathrm{rest}= 231321.8283\, \rm MHz$\footnote{According to the Cologne Database for Molecular Spectroscopy, CDMS, available at \url{https://cdms.astro.uni-koeln.de/}.} have been published by \cite{Sanhueza19} and \cite{Sakai22}, respectively, and we refer to those papers for a complete description of the observations and of the data reduction. Briefly, the data were observed in Cycle 3 (Project ID: 2015.1.01539.S; PI: Sanhueza), with the 12, array, the 7m array {(baselines ranging from 8 to 330$\,$m)}, and the Total Power (the latter for spectral lines only). The data were acquired as mosaics, consisting of 10 pointings for the 12m array and 3 for the 7m array.  \par
{The spectral window containing the \nndp line was imaged using the automatic clean script \textsc{yclean} \citep{Contreras18}, which uses natural weighting and multiscale deconvolver (scales: 0, 3, 10, 30). The \textsc{casa} version 5.4 was used for the imaging. To allow a better comparison with the \olineh data (see Sect. \ref{h2dp_n2dp}), we have excluded the Total Power data for this analysis, and the maximum recoverable scale is $\theta_\mathrm{MRS} =  35''$. The final angular resolution of the 12m+7m combined datacube is $\approx 1''.0\times 1.''4$, and the pectral resolution is  $0.17\, $\kms. The data have been converted from the flux scale to temperature scale through the gain $G =15 \, \rm mK/ (mJy\, beam^{-1}) $, computed with Eq. \ref{gain_eq}. }

\subsection{Band 3 observations}
The Band 3 data were collected as part of project 2018.1.00299.S (PI: Contreras), during Cycle 6. The data consist of 12m array observations (performed in December 2018 and April 2019), 7m array observations (performed in January 2019), and Total Power (April 2019), with baseline ranging from $9.0\, \rm m$ to $500\rm \, m$. The average precipitable water vapour was in the range $4.6\, \mathrm{mm} <PWV < 6\rm \, mm$. The quasars J2000-1748, J1517-2422, and J1832-2039 were used as calibrators for the {12m-array} data, whilst J1751+0939, J2056-4714, and J1911-2006 were used during the {7m-array} observations. \par
In this paper, we focus on the \nnhp (1-0) transition at $93.174\, \rm GHz$, which was targeted by a dedicated SPW {with a spectral resolution of $61\, \rm kHz$, corresponding to a velocity resolution of $0.20\,$\kms at the \nnhp frequency.} The primary beam size at the \nnhp frequency is $\approx 60''$ for the 12m array, and $\approx 110''$ for the 7m array. The line was imaged with Briggs weighting (robust = 0.5) and multiscale deconvolver, using the \texttt{tclean} task of the software \textsc{casa} {(version 5.7).} The scales used were $0,5,15,25$ times the pixel size ($0''.4${, corresponding to 1/4 of the beam minor axis, in agreement with the Nyquist sampling}). The final beam size {of the composite datacube (12m+7m+TP arrays), after primary-beam correction,} is $2''.9\times 1.6''$. {The fluxes where converted in temperature scale using Eq. \ref{gain_eq}, obtaining the gain $G =30 \, \rm mK/ (mJy\, beam^{-1}) $. The maximum recoverable scale considering the 12m and 7m array configuration is $\theta_\mathrm{MRS} = 85''$, but the Total Power observations further increase it.}

\section{Analysis}

\subsection{The prestellar core population\label{CorePop}}
The \olineh emission traces cold and dense gas. In this Section, we describe the analysis of the Band 7 data aimed to identify the population of prestellar cores in the clump, and to study their properties.

\subsubsection{Prestellar cores identification\label{scimes}}
Our aim is to use the \olineh data to identify structures (cores) which are in a early, prestellar stage. Similarly to what {has been done} in \cite{Redaelli21}, we use 
\textsc{scimes} \citep{Colombo15}, which is based on the dendrogram algorithm \citep{Rosolowski08} and analyses data in three-dimensional, position-position-velocity (ppv) space.  \par
\begin{figure}[!h]
\centering
\includegraphics[width=.5\textwidth]{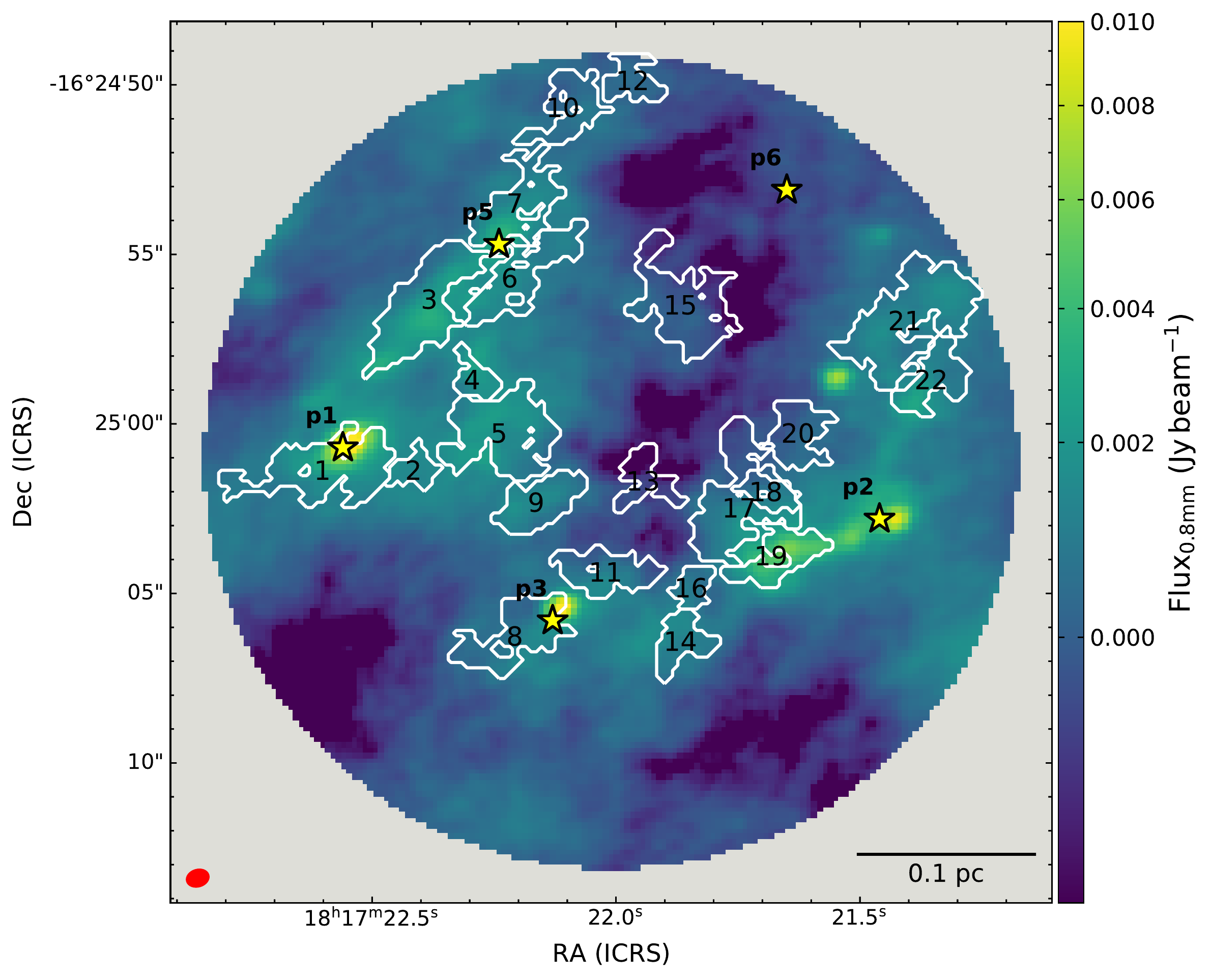}
\caption{ The cores identified in \ohhdp by \textsc{scimes} are shown as white contours, on top of the continuum emission. They are labelled in order of decreasing right-ascension. The star symbols show the position of the protostellar cores identified by \cite{Li20}{, labelled in bold face}. The beam size and scalebar are indicated in the bottom left and right corners, respectively.
\label{cores}}
\end{figure}

The first key step of \textsc{scimes} is the \textit{dilmasking} masking technique, which maximises the information recoverable in low signal-to-noise ratio (S/N) data (see \citealt{Rosolowsky06}). The code identifies regions where the $\rm S/N$ is higher than a given threshold ($\rm S/N_{lim}$), that however contain emission peaks brighter than a second threshold ($\rm S/N_{peak}$). After a few test, we set $\rm S/N_{peak} = 2$, and $\rm S/N_{lim} = 1.5$, consistent with our choice in \cite{Redaelli21}, which maximises the signal recovery. Another key parameter to build the dendrogram is the minimum height (in flux/brightness) that a structure must have to be catalogued as an independent leaf ($\Delta_{min}$). We set the minimum height of an identified structure on $\Delta_{min} = 2.8 \times rms$\footnote{{Values tested in the range $\Delta_{min}=(2.5-3.5)\times rms$ lead to variation in only 18\% of the identified structures. Using $\Delta_{min}/rms=2.8$, instead of 3.0, allows the cores 21 and 22 to be separated, instead of merging is a single structure significantly larger than any other identified.}}, where $rms = 100 \, \rm mK$ (this is the value obtained on the datacube before primary-beam correction, since \textsc{scimes} requires data with constant noise). We set the minimum number of channels that a leaf must span to $N \rm^{min} _{chan} = 2$, and we mask structures smaller than three times the beam area. With these input parameters, we find 22 prestellar cores, shown in Fig. \ref{cores}. Some of them appear to overlap in projection on the plane of the sky. This is due to the fact that \textsc{scimes} works in ppv space, and it is therefore able to identify distinct velocity components as belonging to different structures. We report in Table \ref{AveSpectra_par} the positions and sizes, expressed in terms of effective radius, of the whole sample of cores. \par
Figure \ref{cores} confirms the fact that continuum and \ohhdp emission do not perfectly correlate, as seen also in Fig. \ref{Band7_Data}. The positions of the protostellar candidates found by \cite{Li20}, also shown in Fig. \ref{cores}, are usually associated with peaks of continuum (with the exception of the one in the north-west corner), and either not associated with, or found at the edges of, the \hhdp-identified cores. In Appendix \ref{ContCores} we present a more detailed study of the continuum cores.

\begin{deluxetable*}{cccccccccc}
\tablecaption{Core properties and best-fit results obtained fitting their average spectra with \textsc{mcweeds}. The $rms$ values are standard deviation over line-free channels. Uncertainties on \vlsr, \sigmav, \ncol, one-dimension turbulent Mach number, and virial mass are expressed as 95\% high probability density intervals (HPD).\label{AveSpectra_par}}
\tablehead{
\colhead{Core id}	& \multicolumn2c{Position}	&\colhead{$R_\mathrm{eff}$\tablenotemark{a}}	& \colhead{$rms$}	& \colhead{\vlsr}	& \colhead{\sigmav} 	& \colhead{\ncol}    		& \colhead{$\mathcal{M}$\tablenotemark{b}} & \colhead{$M_\mathrm{vir}$\tablenotemark{b}} \\ 
&\colhead{RA(h:m:s.ss)}& \colhead{Dec (d:m:s.ss)}	&	\colhead{$10^{3}$AU}		& \colhead{K}	& \colhead{\kms}	&\colhead{\kms}		&\colhead{ $\log_{10} ( \rm cm^{-2})$} &      &  \colhead{M$_\odot$}}
\startdata
1	&	$18:17:22.59$	&	$-16:25:1.35$	&	5.5	&	0.09	&	$38.26^{+0.06}_{-0.06}$	&	$0.38^{+0.07}_{-0.06}$	&	$13.20^{+0.06}_{-0.06}$	&	$1.84^{+0.42}_{-0.38}$	&	$4.8^{+1.7}_{-1.4}$	\\ 
2	&	$18:17:22.41$	&	$-16:25:01.36$	&	2.4	&	0.07	&	$38.96^{+0.10}_{-0.11}$	&	$0.34^{+0.02}_{-0.02}$	&	$12.91^{+0.07}_{-0.07}$	&	$1.61^{+0.02}_{-0.02}$	&	$1.74^{+0.04}_{-0.04}$	\\  
3	&	$18:17:22.38$	&	$-16:24:56.44$	&	6.1	&	0.07	&	$40.33^{+0.04}_{-0.05}$	&	$0.40^{+0.02}_{-0.04}$	&	$13.27^{+0.04}_{-0.04}$	&	$2.01^{+0.11}_{-0.24}$	&	$6.1^{+0.6}_{-1.1}$	\\ 
4	&	$18:17:22.29$	&	$-16:24:58.68$	&	2.5	&	0.07	&	$38.38^{+0.05}_{-0.05}$	&	$0.29^{+0.05}_{-0.05}$	&	$13.02^{+0.06}_{-0.06}$	&	$1.32^{+0.34}_{-0.32}$	&	$1.4^{+0.5}_{-0.4}$	\\ 
5	&	$18:17:22.24$	&	$-16:25:0.27$	&	5.5	&	0.03	&	$38.43^{+0.04}_{-0.04}$	&	$0.40^{+0.04}_{-0.03}$	&	$13.04^{+0.03}_{-0.03}$	&	$2.00^{+0.22}_{-0.19}$	&	$5.5^{+1.0}_{-0.9}$	\\ 
6	&	$18:17:22.22$	&	$-16:24:55.66$	&	5.1	&	0.04	&	$39.71^{+0.05}_{-0.05}$	&	$0.30^{+0.01}_{-0.01}$	&	$12.96^{+0.04}_{-0.04}$	&	$1.37^{+0.01}_{-0.01}$	&	$2.96^{+0.03}_{-0.03}$	\\ 
7	&	$18:17:22.21$	&	$-16:24:53.65$	&	4.5	&	0.06	&	$40.42^{+0.04}_{-0.04}$	&	$0.36^{+0.04}_{-0.03}$	&	$13.16^{+0.04}_{-0.04}$	&	$1.74^{+0.22}_{-0.19}$	&	$3.6^{+0.7}_{-0.5}$	\\ 
8	&	$18:17:22.21$	&	$-16:25:6.23$	&	4.5	&	0.06	&	$41.01^{+0.12}_{-0.13}$	&	$0.42^{+0.09}_{-0.09}$	&	$12.72^{+0.11}_{-0.10}$	&	$2.08^{+0.50}_{-0.52}$	&	$4.8^{+2.1}_{-1.7}$	\\ 
9	&	$18:17:22.16$	&	$-16:25:2.31$	&	3.9	&	0.04	&	$39.23^{+0.03}_{-0.03}$	&	$0.30^{+0.03}_{-0.03}$	&	$12.96^{+0.04}_{-0.04}$	&	$1.38^{+0.20}_{-0.20}$	&	$2.3^{+0.4}_{-0.4}$	\\ 
10	&	$18:17:22.10$	&	$-16:24:50.74$	&	3.7	&	0.21	&	$40.43^{+0.11}_{-0.12}$	&	$0.33^{+0.10}_{-0.09}$	&	$13.26^{+0.11}_{-0.14}$	&	$1.58^{+0.59}_{-0.55}$	&	$2.6^{+1.5}_{-1.1}$	\\ 
11	&	$18:17:22.02$	&	$-16:25:4.36$	&	3.9	&	0.04	&	$41.21^{+0.06}_{-0.06}$	&	$0.33^{+0.01}_{-0.02}$	&	$12.76^{+0.05}_{-0.06}$	&	$1.60^{+0.04}_{-0.04}$	&	$2.7^{+0.1}_{-0.1}$	\\ 
12	&	$18:17:21.97$	&	$-16:24:49.84$	&	2.9	&	0.33	&	$40.57^{+0.15}_{-0.14}$	&	$0.28^{+0.02}_{-0.05}$	&	$13.34^{+0.16}_{-0.16}$	&	$1.26^{+0.12}_{-0.12}$	&	$1.5^{+0.2}_{-0.2}$	\\ 
13	&	$18:17:21.94$	&	$-16:25:1.57$	&	3.1	&	0.04	&	$38.86^{+0.05}_{-0.05}$	&	$0.21^{+0.04}_{-0.04}$	&	$12.50^{+0.08}_{-0.08}$	&	$0.83^{+0.29}_{-0.30}$	&	$1.0^{+0.3}_{-0.3}$	\\ 
14	&	$18:17:21.87$	&	$-16:25:6.43$	&	3.2	&	0.07	&	$41.46^{+0.06}_{-0.05}$	&	$0.34^{+0.05}_{-0.05}$	&	$13.11^{+0.06}_{-0.06}$	&	$1.64^{+0.29}_{-0.28}$	&	$2.4^{+0.7}_{-0.6}$	\\ 
15	&	$18:17:21.86$	&	$-16:24:56.45$	&	5.3	&	0.04	&	$39.55^{+0.04}_{-0.04}$	&	$0.22^{+0.03}_{-0.03}$	&	$12.76^{+0.06}_{-0.06}$	&	$0.91^{+0.21}_{-0.23}$	&	$1.9^{+0.4}_{-0.4}$	\\ 
16	&	$18:17:21.84$	&	$-16:25:4.75$	&	2.4	&	0.08	&	$41.44^{+0.04}_{-0.04}$	&	$0.29^{+0.04}_{-0.04}$	&	$13.15^{+0.05}_{-0.05}$	&	$1.28^{+0.09}_{-0.18}$	&	$1.2^{+0.1}_{-0.2}$	\\ 
17	&	$18:17:21.75$	&	$-16:25:2.60$	&	6.2	&	0.03	&	$39.28^{+0.03}_{-0.03}$	&	$0.24^{+0.03}_{-0.03}$	&	$12.77^{+0.04}_{-0.05}$	&	$1.02^{+0.19}_{-0.21}$	&	$2.5^{+0.5}_{-0.5}$	\\ 
18	&	$18:17:21.68$	&	$-16:25:1.99$	&	2.7	&	0.06	&	$37.93^{+0.06}_{-0.06}$	&	$0.19^{+0.08}_{-0.06}$	&	$12.61^{+0.11}_{-0.12}$	&	$0.66^{+0.56}_{-0.49}$	&	$0.8^{+0.6}_{-0.3}$	\\ 
19	&	$18:17:21.67$	&	$-16:25:3.87$	&	4.2	&	0.05	&	$41.34^{+0.02}_{-0.02}$	&	$0.27^{+0.02}_{-0.02}$	&	$13.20^{+0.03}_{-0.03}$	&	$1.25^{+0.13}_{-0.13}$	&	$2.2^{+0.3}_{-0.3}$	\\ 
20	&	$18:17:21.63$	&	$-16:25:0.29$	&	3.7	&	0.06	&	$39.28^{+0.07}_{-0.07}$	&	$0.31^{+0.06}_{-0.06}$	&	$12.82^{+0.07}_{-0.08}$	&	$1.49^{+0.36}_{-0.34}$	&	$2.3^{+0.8}_{-0.7}$	\\ 
21	&	$18:17:21.40$	&	$-16:24:56.93$	&	6.2	&	0.11	&	$40.25^{+0.04}_{-0.04}$	&	$0.29^{+0.03}_{-0.03}$	&	$13.35^{+0.05}_{-0.05}$	&	$1.35^{+0.20}_{-0.20}$	&	$3.5^{+0.7}_{-0.7}$	\\ 
22	&	$18:17:21.35$	&	$-16:24:58.56$	&	3.6	&	0.14	&	$40.36^{+0.06}_{-0.06}$	&	$0.30^{+0.04}_{-0.05}$	&	$13.29^{+0.07}_{-0.08}$	&	$1.42^{+0.26}_{-0.28}$	&	$2.2^{+0.6}_{-0.5}$\\
\enddata
\tablenotetext{a}{The effective radius is the radius of a circular region with the same area of the core.}
\tablenotetext{b}{The one-dimensional turbulent Mach number and the virial masses are computed assuming $T_\mathrm{gas} = 10 \, \rm K$.}		
\end{deluxetable*}

\subsubsection{Core properties from \ohhdp fitting\label{mcweeds}}
\begin{figure*}
\centering
\includegraphics[width=\textwidth]{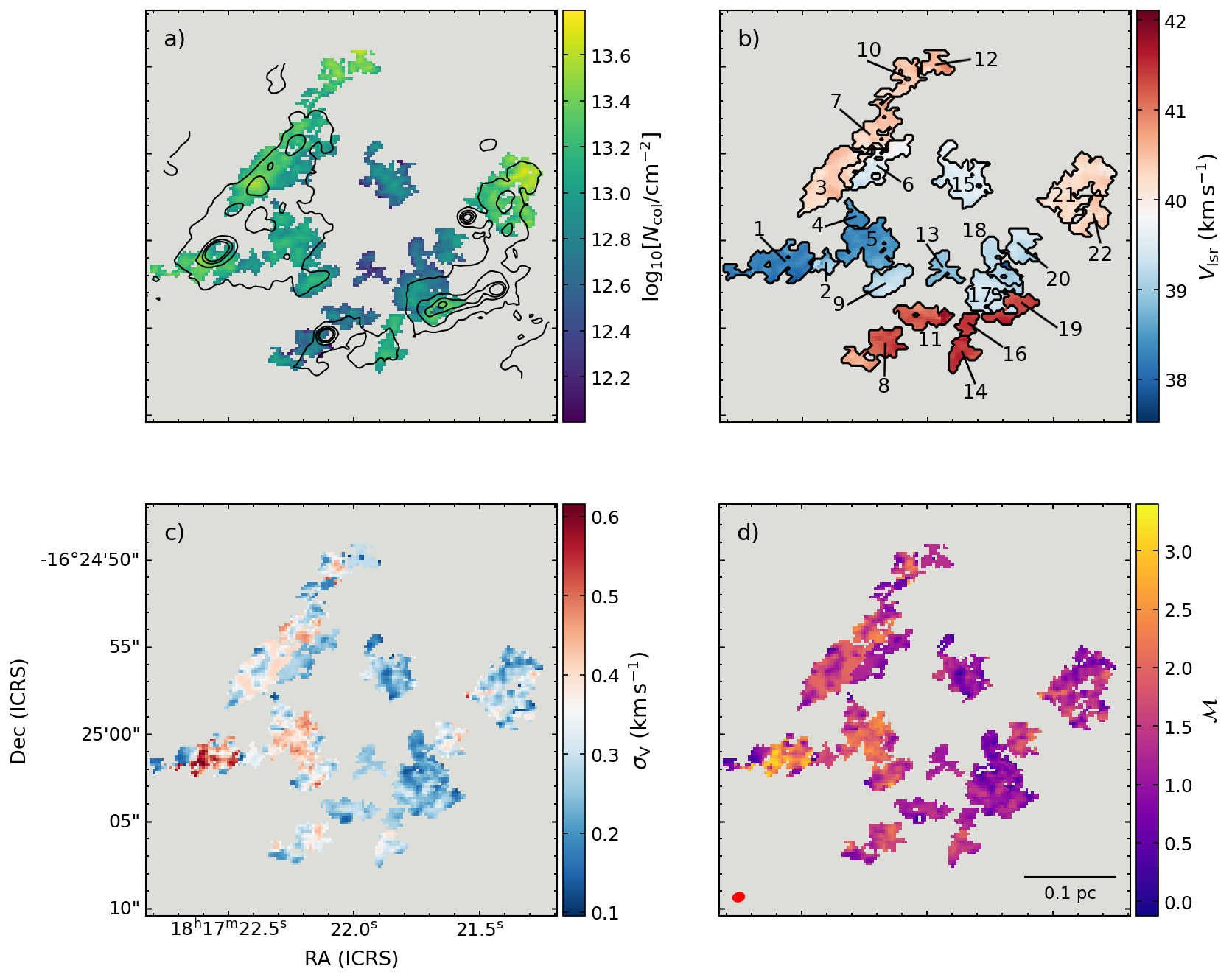}
\caption{Composite maps obtained showing together the results of \textsc{mcweeds} fit on the \ohhdp spectra for all the cores. \textit{Panel a)}: molecular column density, with contours of the $0.8\, \rm mm$ continuum emission overlaid (levels from $2$ to $11\sigma$, in steps of $3\sigma$); \textit{Panel b)}: \vlsr map, with the \ohhdp-identified cores shown in black contours, and numeric labels; \textit{Panel c)}: velocity dispersion \sigmav map; \textit{Panel d)}: one-dimensional turbulent Mach number. Note that in case of core overlapping, the maps show the results of only one of them. For the complete sets of maps of each individual core we refer to Appendix. \ref{AllCoresMaps}. The beam size and scalebar are indicated in the bottom left and right corners, respectively, of panel d).
\label{MCweeds_Results}}
\end{figure*}

We perform a spectral fit of the \olineh in each core, in order to derive maps of the centroid velocity (\vlsr), linewidth ($FWHM$), and column density $N_\mathrm{col} (\text{\ohhdp})$. We use the parallelised version of the \textsc{mcweeds} code \citep{Giannetti17}, which is based on the \textsc{Weeds} package of \textsc{GILDAS} \citep{Maret11}. {\textsc{Weeds} is able to produce synthetic spectra in LTE approximation based on a set of input parameters ($FWHM$, \vlsr, molecular column density, excitation temperature, and source size), assuming that the line profile is Gaussian. \textsc{mcweeds}, instead, provides the framework to optimise the search for the best-fit solution of the parameters\footnote{The source size is selected to ensure that the beam filling factor is unity.}.} The code analyses the spectrum in each pixel with Bayesian statistical models implemented using \textsc{PyMC} \citep{Patil10}. In particular, we use a Markov chain Monte Carlo (MCMC) algorithm to sample the parameter space, with uninformative flat priors over the models' free parameters. Similarly as in \cite{Redaelli21}, for each position the code performs 100000 iterations, with a burn-in of 1000 steps. For the excitation temperature, we assume $\text{\tex} = 10 \, \rm K$ (see for instance \citealt{Caselli08, Friesen14, Redaelli21}). The initial guesses for the free parameters are selected individually for each core. \textsc{mcweeds} uses the line $FWHM$ as a free parameter, but here we show the velocity dispersion instead ($\sigma_\mathrm{V} = FWHM/(2 \sqrt{2\, \rm ln(2)}$).
Figure \ref{MCweeds_Results} shows the best-fit maps of the free parameters, obtained composing together those of the single cores. We show the best-fit parameter maps for each core individually in Appendix \ref{AllCoresMaps}.

\par
\begin{figure*}
\centering
\includegraphics[width=\textwidth]{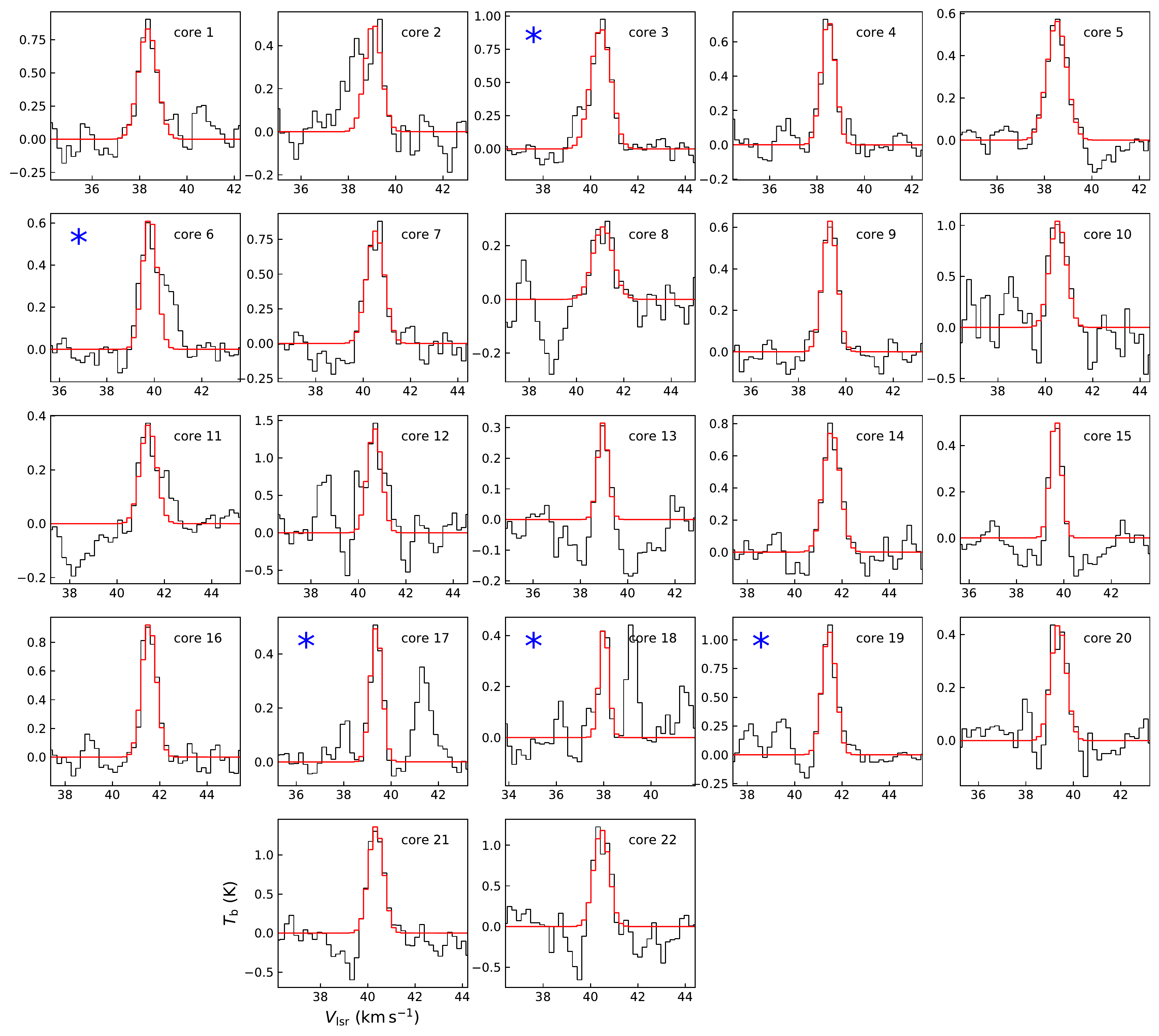}
\caption{In black we show the average \olineh spectra obtained in each core (labelled at the top-right corner of each panel). The red histogram shows the best-fit obtained with \textsc{mcweeds}. The corresponding best-fit parameters are recorded in Table \ref{AveSpectra_par}. {Cores that overlap by at least 5\% of their extension with another core are marked with a blue asterisk in the top-left corner of the panel.}
\label{AveSpectra_h2dp}}
\end{figure*}

The centroid velocity shows little gradient within each core. Excluding core 6 (one of the largest in terms of physical size) and core 12, the dispersion of \vlsr around the average is less than $0.20\,$\kms. However, {a clear change in \vlsr of the order of $3-4 \,$\kms is} visible at the clump level, in particular with changing declination. We can identify three separate groups: i) the southernmost cores have typical velocities of $>41 \, $\kms; ii) the cores in the central part of the clump present lower velocities (\vlsr$< 40 \,$\kms). Group 1 and 2 overlap in the west part of the clump (see e.g. core 19, 17, and 18); iii) in the northern part of AG14 the cores have typical velocities of $40-40.5\,$\kms. The presence of these three sub-populations of cores, with distinct velocities, suggests that AG14 presents a complex kinematics, with several velocity components that spatially overlap (see also the average spectra in Fig.~\ref{AveSpectra_h2dp}). This is further investigated in Sect. \ref{kinematics}.
\par
\begin{figure}[!t]
\centering
\includegraphics[width=.48\textwidth]{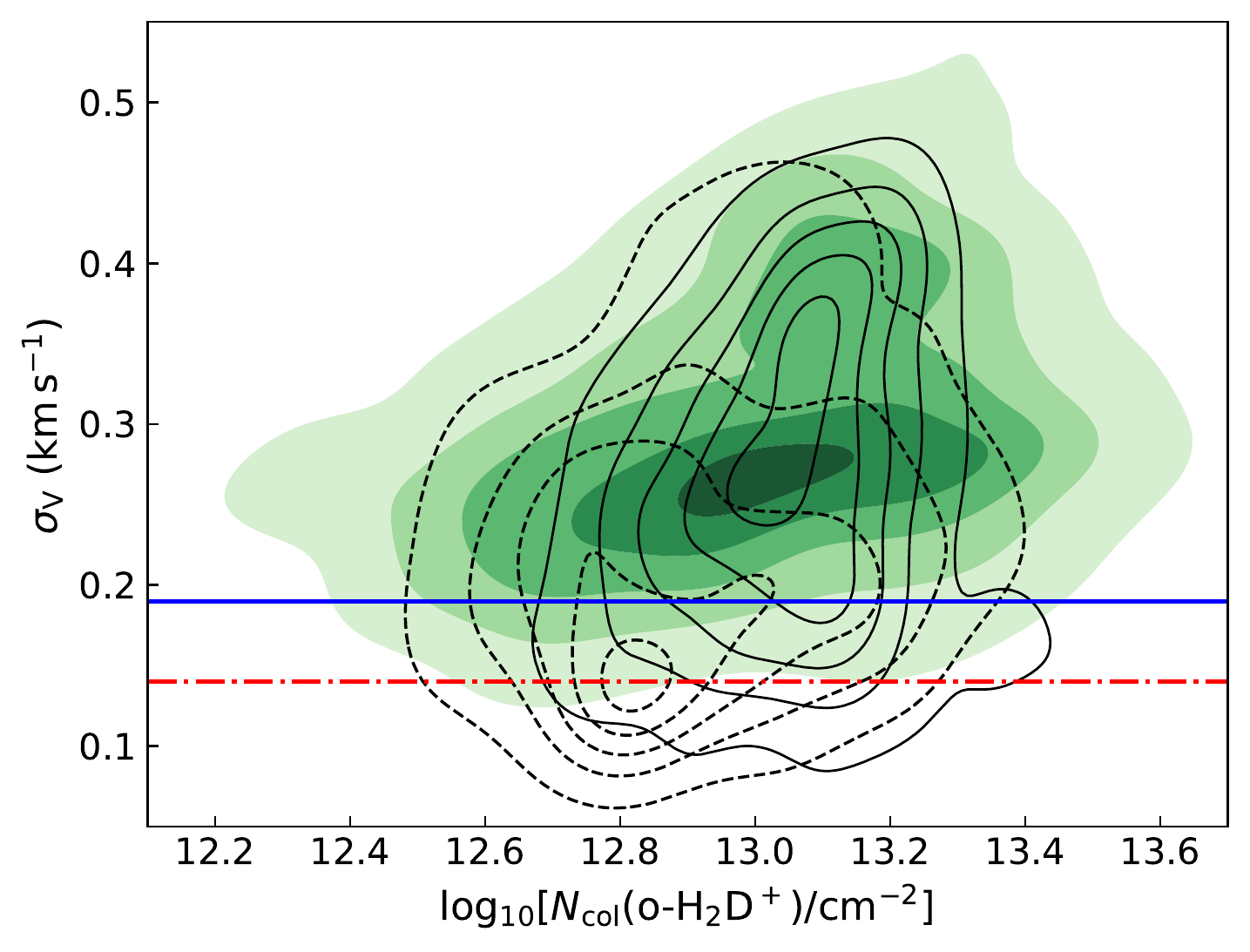}
\caption{The green colorscale shows the normalised kernel density distribution of \sigmav and \ncol in AG14, with contour levels set on  $[0.1, 0.3, 0.5, 0.7, 0.9]$. The same data are shown for AG351 with the dashed contours, and for AG354 with solid contours (taken from \citealt{Redaelli21}; the \ncol values have been corrected by a factor $\sqrt{\pi}$, see text for details). The horizontal blue solid and red dashed-dotted lines represent the sound speed and the thermal broadening of \ohhdp, respectively, both at $10\, $K.
\label{DensityPlot}}
\end{figure}

To investigate the core properties in terms of velocity dispersion and column density, we present the density distribution of these two parameters in Fig. \ref{DensityPlot} (green colorscale), and compare it with the results for AG351 and AG354 obtained in \cite{Redaelli21}\footnote{Due to a typo in the code, the column density values of \cite{Redaelli21} where overestimated by a factor of $\sqrt{\pi}$. This does not affect the trends found in the comparison between the sources. However, in order to compare the results in AG14 with the ones in the other two sources, we corrected the latter before producing the plot in Fig. \ref{DensityPlot}}. {We highlight that AG351 and AG354 are at about half of the distance with respect to AG14. However, the Band 7 data (both continuum and lines) were acquired with a higher angular resolution for AG14 ($\approx 0''.55$, to be compared with $\approx 0''.9$ for AG351 and AG354). Hence, the linear resolution of the data is only $\approx 25$\% worse, allowing a fair comparison.} The average {\ohhdp} column density in AG14 is $\langle N_\mathrm{col}\rangle = 10^{13} \, \rm cm^{-2}$, which is consistent with the value obtained by \cite{Sabatini20} with  observations from the Atacama Pathfinder EXperiment (APEX), at a resolution ($17''$) comparable to the FoV of the ALMA data. The average velocity dispersion is $\langle \sigma_\mathrm{V} \rangle = 0.30\,$\kms. AG14 presents on average higher column density values than AG351. 
Furthermore, {both} clumps reported in \cite{Redaelli21} showed very narrow lines, with a significant fraction on positions below both the isothermal sound speed at $10\, \rm K$ ($c_\mathrm{s} = 0.19\, $\kms, assuming a gas mean molecular weight $\mu = 2.33$) and the thermal broadening of the \olineh at $ 10\, \rm K$ ($\sigma_\mathrm{V, th} =0.14\, $\kms). On the contrary, in AG14 only 8\% of the positions detected in \ohhdp present $\sigma_V < c_\mathrm{s} $ (to be compared with 36\% and 23\% in AG351 and AG354, respectively), and less than 1\% are characterised by $ \sigma_\mathrm{V} <\sigma_\mathrm{V, th}$ (17\% and 7\% in AG351 and AG354). {The gas motions in AG14 hence appear less quiescent than in AG351 and AG354.} We highlight that the derived velocity dispersion values might be overestimated, due to the limited spectral resolution of our observations. Lines narrower than $FWHM = 0.6 \, $\kms (corresponding to $\sigma_\mathrm{V} = 0.25\,$\kms), in fact, are resolved by less than three channels. However, the spectral resolution is the same for all three clumps, and therefore {this problem would not affect the comparison between the sources.} {In Appendix \ref{missingFlux} we discuss also the linewidth overestimation due to opacity effect, which is found to be at most 15\% and only in the densest parts of the AG14.}

\par
From the total velocity dispersion \sigmav, the non-thermal contribution can be computed,  in the assumption that the thermal and non-thermal contributions are independent and thus they sum in quadrature \citep[see e.g.][]{Myers91}:
\begin{equation}
\sigma_\mathrm{V,NT}  = \sqrt{ \sigma_\mathrm{V}^2  - \sigma_\mathrm{V, th} ^2 }  = \sqrt{ \sigma_\mathrm{V}^2 - \frac{k_\mathrm{B}T_\mathrm{gas}}{m_\mathrm{H_2D^+}} } \; ,\label{Mach}
\end{equation}
where  $m_\mathrm{H_2D^+}$ is the \hhdp molecular mass (in g, 4 a.m.u.), $T_\mathrm{gas}$ is the gas temperature (assumed to be $10 \, \rm K$), and $k_\mathrm{B}$ is the Boltzmann constant.  The one-dimensional turbulent Mach number is then $ \mathcal{M} = \sigma_\mathrm{V,NT} /c \rm _s$. The bottom-right panel of Fig.~\ref{MCweeds_Results} shows the map of this parameter. In most of the cores, the turbulent motions are transonic or mildly supersonic ($\mathcal{M} = 1-2$). A few cores, however, present subsonic non-thermal linewidhts (e.g. cores n. 15, 17, 18, and 19).

\subsubsection{Average core properties}
An assessment of the dynamical state of each core can be obtained from the one-dimensional turbulent Mach number and the virial mass. These quantities are computed by fitting the averaged spectra within each core via \textsc{mcweeds}. The average spectra in each core, together with the obtained best-fit models, are shown in Fig.~\ref{AveSpectra_h2dp}, and the best-fit values are presented in Table \ref{AveSpectra_par}. {In Fig. \ref{AveSpectra_h2dp} we have highlighted with a blue asterisk those cores with significant overlap (at least 5\% of their extension) with at least one other core. These cases present either multiple velocity components well separated in velocity (e.g. core 17 and 18), or broad wings and shoulders (core 6). We however select the initial guesses for the fit of the average spectra from the results of the pixel-by-pixel fit of each core, and \textsc{mcweeds} is hence able to identify and fit the correct velocity component.} \par
From the \sigmav values derived fitting the average spectra we computed the one-dimensional turbulent Mach number in each core, following the procedure described in Sect. \ref{mcweeds}. Furthermore, we have derived the total velocity dispersion of the gas ($\sigma_\mathrm{dyn}$) as:
\begin{equation}
\sigma_\mathrm{dyn} = \sqrt{\sigma_\mathrm{NT}^2 + c_\mathrm{s}^2 } \; ,
\end{equation}
from which we can derive the virial mass of the cores using the equation of \cite{Bertoldi92}, in the assumption of uniform density \citep{MacLaren88}:
\begin{equation}
M_\mathrm{vir} = \frac{5 R_\mathrm{core} \sigma_\mathrm{dyn}^2 }{G} =  1200 \times \left ( \frac{R_\mathrm{core}}{\mathrm{pc}} \right ) \left ( \frac{\sigma_\mathrm{dyn}}{\mathrm{km \, s^{-1}}} \right )^2 \rm{M_\odot} \; ,
\label{Mvir}
\end{equation}
where for the $R_\mathrm{core}$ values we have used the effective radii listed in Table \ref{AveSpectra_par}. This definition of the virial mass ignores contributes from magnetic fields and external pressure. \par
The values of $\mathcal{M}$ span the range $0.7- 2.0$, with an average of $\langle \sigma_\mathrm{V,NT} /c \rm _s\rangle = 1.4$. The turbulent motions in AG14 are transonic, or mildly supersonic. The $\sigma_\mathrm{V,NT} /c \rm _s$ values are significantly lower than the value reported by \cite{Sabatini20} using APEX observations ($\sigma_\mathrm{V,NT} /c \rm _s = 5.4$), most likely because the unresolved single-dish spectrum overestimates the linewidth, due to the velocity gradient that the ALMA data unveils ($\approx 4 \,$\kms) and the presence of several velocity components. The virial masses derived at $10 \, \rm K$ are found within the range $0.8-6.1 \, \rm M_\odot$, with 50\% of the cores presenting $M\rm _{vir} < 2.4 \, M_\odot $. If the prestellar cores identified in \ohhdp are virialised, they are essentially low-mass. 
\par
In the analysis of the \olineh transition performed so far, we have assumed that the line opacity is low, and that the missing flux due to the filtering of the large scale emission from the interferometer is negligible. We discuss these points in further detail in Appendix ~\ref{missingFlux}.

\subsubsection{Continuum emission \label{contBand7}}
Further information on the core properties comes from the analysis of the continuum emission. In particular, we can estimate the core total mass (\mcore) using the equation:
\begin{equation}
M_\mathrm{core} = f \frac{D^2 S_\mathrm{tot}}{B_\nu(T_\mathrm{dust}) \kappa_\nu } \; ,
\label{Mdust}
\end{equation}
where $f$ is the gas-to-dust ratio (assumed to be 100, \citealt{Hildebrand83}); $D$ is the source's distance; $B_\nu(T_\mathrm{dust})$ is the Planck function at the frequency $\nu = 371 \, \rm GHz$ and temperature $T_\mathrm{dust}$; $S_\mathrm{tot}$ is the $0.8\, \rm mm$ total flux integrated within the contours of the \ohhdp-identified cores, and $ \kappa_\nu$ is the dust opacity at the frequency of the observations. For the latter, we use the power-law expression:
\begin{equation}
\kappa_\nu = \kappa_0 \left( \frac{\nu}{\nu_0}\right)^\beta = 1.71 \, \rm  cm^2 \, g^{-1} \; , 
\label{kappa}
\end{equation}
in which we use $\beta = 1.5$ for the dust emissivity index \citep{Mezger90,Walker90} and $\kappa_0 =10\, \rm  cm^2 \, g^{-1}$ for the dust opacity at the reference wavelength $\lambda_0 = 250 \, \mu \rm m$ \citep{Hildebrand83,Beckwith90}. In the assumption of spherical symmetry and uniform gas distribution, we can evaluate the gas density as:
\begin{equation}
 {n\mathrm{(H_2)} =  \frac{ 3 M_\mathrm{core}}{ 4 \pi R_\mathrm{eff}^3 \mu_\mathrm{H_2} m_\mathrm{H} } \; ,}
 \label{nh2}
\end{equation}
where $m_\mathrm{H}$ and $\mu_\mathrm{H_2} = 2.8$ are respectively the hydrogen mass and the gas mean molecular
weight per hydrogen molecule \citep{Kauffmann08}. \par

\begin{deluxetable*}{c|ccc|ccc|ccc}
\tablecaption{Core properties derived from the continuum emission at $0.8\, \rm mm$. The core masses, volume densities, and virial parameters are computed at three distinct temperatures: 10, 15, and 20$\,$K. The uncertainty on core masses and densities is 38\%, whilst it is 43\% on the virial parameter values. \label{CoreProp2}}
\tablehead{\colhead{Core id}	& \colhead{$M_\mathrm{core}$ }& \colhead{$n \rm (H_2)$} &	\colhead{$\alpha_\mathrm{vir}$}	& \colhead{$M_\mathrm{core}$} & \colhead{$n \rm (H_2)$} &	\colhead{$\alpha_\mathrm{vir}$	}& \colhead{$M_\mathrm{core}$} & \colhead{$n \rm (H_2)$}    & \colhead{$\alpha_\mathrm{vir}$} \\ 
\colhead{}		&\colhead{$\mathrm{M_\odot}$}	&	\colhead{$10^6 \rm cm^{-3}$}& \colhead{}	&\colhead{$\mathrm{M_\odot}$}	&	\colhead{$10^6 \rm cm^{-3}$}	&\colhead{}	&\colhead{$\mathrm{M_\odot}$}	&	\colhead{$10^6 \rm cm^{-3}$}	 & \colhead{}}
\startdata
 & \multicolumn{3}{c|}{$10 \, \rm K$} & \multicolumn{3}{c|}{$15 \, \rm K$} & \multicolumn{3}{c}{$20 \, \rm K$}  \\
\hline 
1	&	$27 \pm 10$ &	$4.9 \pm 1.9$ &$0.18\pm0.08$	&	$13 \pm 5$ &	$2.3 \pm 0.9$ &$0.40\pm0.17$	&	$8 \pm 3$ &	$1.4 \pm 0.5$ &$0.7\pm0.3$	\\ 
2	&	$1.9 \pm 0.7$ &	$4.0 \pm 1.5$ &$0.9\pm0.4$	&	$0.9 \pm 0.3$ &	$1.8 \pm 0.7$ &$2.1\pm0.9$	&	$0.5 \pm 0.2$ &	$1.2 \pm 0.4$ &$3.6\pm1.5$	\\ 
3	&	$19 \pm 7$ &	$2.5 \pm 0.9$ &$0.33\pm0.14$	&	$9 \pm 3$ &	$1.1 \pm 0.4$ &$0.8\pm0.3$	&	$5 \pm 2$ &	$0.7 \pm 0.3$ &$1.2\pm0.5$	\\ 
4	&	$2.2 \pm 0.8$ &	$4.3 \pm 1.6$ &$0.6\pm0.3$	&	$1.0 \pm 0.4$ &	$2.0 \pm 0.8$ &$1.5\pm0.6$	&	$0.6 \pm 0.2$ &	$1.2 \pm 0.5$ &$2.5\pm1.1$	\\ 
5	&	$9 \pm 3$ &	$1.5 \pm 0.6$ &$0.7\pm0.3$	&	$3.9 \pm 1.5$ &	$0.7 \pm 0.3$ &$1.5\pm0.6$	&	$2.5 \pm 0.9$ &	$0.4 \pm 0.2$ &$2.4\pm1.0$	\\ 
6	&	$10 \pm 4$ &	$2.1 \pm 0.8$ &$0.31\pm0.13$	&	$4.4 \pm 1.7$ &	$1.0 \pm 0.4$ &$0.7\pm0.3$	&	$2.8 \pm 1.0$ &	$0.6 \pm 0.2$ &$1.2\pm0.5$	\\ 
7	&	$9 \pm 4$ &	$3.1 \pm 1.2$ &$0.39\pm0.17$	&	$4.3 \pm 1.6$ &	$1.5 \pm 0.6$ &$0.9\pm0.4$	&	$2.7 \pm 1.0$ &	$0.9 \pm 0.3$ &$1.5\pm0.6$	\\ 
8	&	$6 \pm 2$ &	$1.9 \pm 0.7$ &$0.8\pm0.4$	&	$2.7 \pm 1.0$ &	$0.9 \pm 0.3$ &$1.9\pm0.8$	&	$1.7 \pm 0.6$ &	$0.6 \pm 0.2$ &$3.1\pm1.3$	\\ 
10	&	$2.8 \pm 1.1$ &	$1.6 \pm 0.6$ &$0.9\pm0.4$	&	$1.3 \pm 0.5$ &	$0.8 \pm 0.3$ &$2.2\pm0.9$	&	$0.8 \pm 0.3$ &	$0.5 \pm 0.2$ &$3.6\pm1.6$	\\ 
17	&	$8 \pm 3$ &	$1.0 \pm 0.4$ &$0.32\pm0.14$	&	$3.7 \pm 1.4$ &	$0.5 \pm 0.2$ &$0.8\pm0.3$	&	$2.3 \pm 0.9$ &	$0.3 \pm 0.1$ &$1.3\pm0.6$	\\ 
18	&	$1.0 \pm 0.4$ &	$1.6 \pm 0.6$ &$0.8\pm0.3$	&	$0.5 \pm 0.2$ &	$0.8 \pm 0.3$ &$1.9\pm0.8$	&	$0.30 \pm 0.11$ &	$0.5 \pm 0.2$ &$3.3\pm1.4$	\\ 
19	&	$13 \pm 5$ &	$5 \pm 2$ &$0.16\pm0.07$	&	$6 \pm 2$ &	$2.5 \pm 0.9$ &$0.38\pm0.16$	&	$3.9 \pm 1.5$ &	$1.6 \pm 0.6$ &$0.6\pm0.3$	\\ 
21	&	$20 \pm 7$ &	$2.4 \pm 0.9$ &$0.18\pm0.08$	&	$9 \pm 3$ &	$1.1 \pm 0.4$ &$0.42\pm0.18$	&	$6 \pm 2$ &	$0.7 \pm 0.3$ &$0.7\pm0.3$	\\ 
22	&	$8 \pm 3$ &	$5.0 \pm 1.9$ &$0.28\pm0.12$	&	$3.6 \pm 1.4$ &	$2.3 \pm 0.9$ &$0.6\pm0.3$	&	$2.3 \pm 0.9$ &	$1.5 \pm 0.6$ &$1.1\pm0.5$\\
\enddata	
\end{deluxetable*}

Equation \ref{Mdust} and, as a consequence, Eq. \ref{nh2} depend on the dust temperature. In the hypothesis that the line is excited in LTE conditions (which holds for $n(\mathrm{H_2}) \gtrsim n_\mathrm{cr} \approx 10^5 \, \rm cm^{-3}$, \citealt{Hugo09}) and that the gas and dust are thermally coupled (which requires $n(\mathrm{H_2}) \gtrsim 10^{4-5}\rm \, cm^{-3}$, \citealt{Goldsmith01}), we can assume that $T_\mathrm{dust} = T_\mathrm{gas} = T_\mathrm{ex}(\text{\ohhdp}) = 10\,$K. However, in order to relax these assumptions and to take into consideration that locally the dust and gas temperatures could differ, we have computed the core masses and average densities at three temperatures, equal to 10, 15, and 20$\,$K. The obtained values are summarised in Table \ref{CoreProp2}. From this analysis, we exclude cores that are undetected in continuum, {meaning that they} lack of peak flux above the $3\sigma$ level. At $10\, \rm K$, the point-like mass sensitivity of our observations is $0.6\, \rm M_\odot$ ($3\sigma$ level). Due to the different morphology that the continuum and molecular line data present, as discussed in Sect. \ref{scimes}, eight cores are excluded.\par
{Regarding uncertainties, we follow \cite{Sanhueza17} {(see in particular their Sect. 5.6)}, and we assume} a 23\% uncertainty on the dust-to-mass-ratio, and 28\% uncertainty on the dust opacity. Furthermore, we assume a 10\% uncertainty on the source's distance. Hence, the uncertainty on the mass and on the density values are $38$\%. {In Equation \ref{Mdust}, the total flux is computed integrating the continuum data within each core masks; in case of core overlap, naturally, this will cause an overestimation of their masses. This problem is more severe with increasing overlap area. We have estimated the significance of this bias using the method presented by \cite{Li20b} to decompose the dust-estimated masses of cores when spectroscopic data are available, in the hypothesis that the molecular transition is a high-density tracer (i.e. it traces densities higher than the threshold for dust-gas coupling) and is optically thin. Under these assumptions, which are both reasonably valid for our \ohhdp data, one can decompose the continuum flux into different cores according to the ratio of the line integrated intensity of each velocity component with respect to the total integrated intensity (computed over all the velocity components). We have performed this analysis for the five cores that overlap by more than 5\% of their area (see also Fig. \ref{AveSpectra_h2dp}). We find that on average their masses are overestimated by 23\%, i.e. less than the uncertainties here considered. We conclude that this possible bias does not affect significantly our results.}
\par
Focusing on the gas density values, we note that also assuming a higher dust temperature of 20$\,$K, all the cores have average densities higher than $3 \times 10^5 \, \rm cm^{-3}$. This level is comparable to the critical density of the \olineh line, corroborating the assumption both of LTE conditions for this transition, and of dust-gas coupling. Regarding the masses, all cores are less massive than $30 \, \rm M_\odot$ at any temperature value here considered. This is consistent with what is found by \cite{Sanhueza19}, who identified cores (at any evolutionary stage) in continuum Band 6 data.  We however highlight that the lack of Total Power observations in continuum can lead to partial filter-out of the large scale emission, hence underestimating the mass values\footnote{The integrated flux over the ALMA Band 7 FoV computed from the APEX $870\, \rm \mu m$ (from the ATLASGAL survey) is $\approx 2.5\, \rm Jy$, whilst the total integrated flux in the ALMA data is only $0.8 \, \rm Jy$, suggesting a significant loss of flux in the large scale emission.}.\par
In Table \ref{CoreProp2} we report also the virial parameter values ($\alpha_\mathrm{vir} = M_\mathrm{vir}/ M_\mathrm{core}$) at the three temperatures here considered. The uncertainty on $\alpha_\mathrm{vir}$ takes into account the 38\% uncertainty on the core masses and a further 20\% error, which corresponds to the average uncertainty on the virial mass values. At low dust temperature ($T_\mathrm{dust}=10\,\rm K$), all cores present $\alpha_\mathrm{vir}< 1.0$, suggesting that they are both subvirial and gravitationally bound. Also at $T_\mathrm{dust}=15\,\rm K$, we derived $0.3 \leq \alpha_\mathrm{vir}\leq 2.2$, and all cores are gravitationally bound ($\alpha_\mathrm{vir}< 2$) within uncertainties. The virial parameter increaseas with temperature, but still at $T_\mathrm{dust}=20\,\rm K$ 50\% of the cores in the sample are subvirial within uncertainties. In particular, the most massive cores present the lowest $\alpha_\mathrm{vir}$ values ($\alpha_\mathrm{vir}\lesssim 0.3 $ for $\text{\mcore}\gtrsim 8\rm \, M_\odot$), in agreement with several observational results (see e.g. \citealt{Kauffmann13}). This suggests that the largest cores in the sample are not in equilibrium, unless other sources of pressure (e.g. magnetic fields) contribute to the virialisation.\par
 {\cite{Singh21} performed an extensive study regarding biases in the computation of the virial parameter that tend to lead to its underestimation. Those authors in particular discussed the role of \textit{i)} neglecting the gas bulk motions in the calculation of $\sigma_\mathrm{dyn}$ and of \textit{ii)} the subtraction of the background emission. They found that when these aspects are taken into account, many cores that appeared subvirial become instead virialised or supervirial. However, our analysis intrinsically limits this problem. Since $\sigma_\mathrm{dyn}$ is computed from averaged spectra in each core, bulk motions ---if present--- are already taken into account as they increase the velocity dispersion of the averaged signal. Furthermore, as noted also by \cite{Singh21}, interferometric observations naturally filter out the large scale emission, hence performing an approximate background subtraction. We conclude that these effects are likely negligible in our results. } 
\par
{We now discuss the properties of the most massive cores identified.} Core 1, with $M_\mathrm{core} \approx 30 \, M_\odot$ at $10 \, \rm K$ is the most massive core and it is subvirial at any temperature value considered in this work. However, we have reasons to believe that this core is not in a prestellar stage. In fact, it overlaps with a continuum core associated with outflow emission and protostellar activity (see Appendix \ref{ContCores}). The continuum flux peak is found close to the edge of the core, suggesting that \ohhdp is tracing the part of the envelope which is still cold and dense enough to emit the \olineh transition. {A similar discussion could be made for core 3, which has $M_\mathrm{core}= 20\, \rm M_\odot$ (at 10$\,$K) and it is subvirial, and it lays in close proximity to the protostar p5. Also core 21 has a similar mass, but unlike the other two, no protostellar core appears to be found in its surroundings. However, a significant continuum peak is found just outside its south-east edge. This peak is associated with the continuum-identified structure c8 (see Appendix \ref{ContCores}, where we speculated about the evolutionary stage of this core). }
\par
\begin{figure}[!h]
\centering
\includegraphics[width=.5\textwidth]{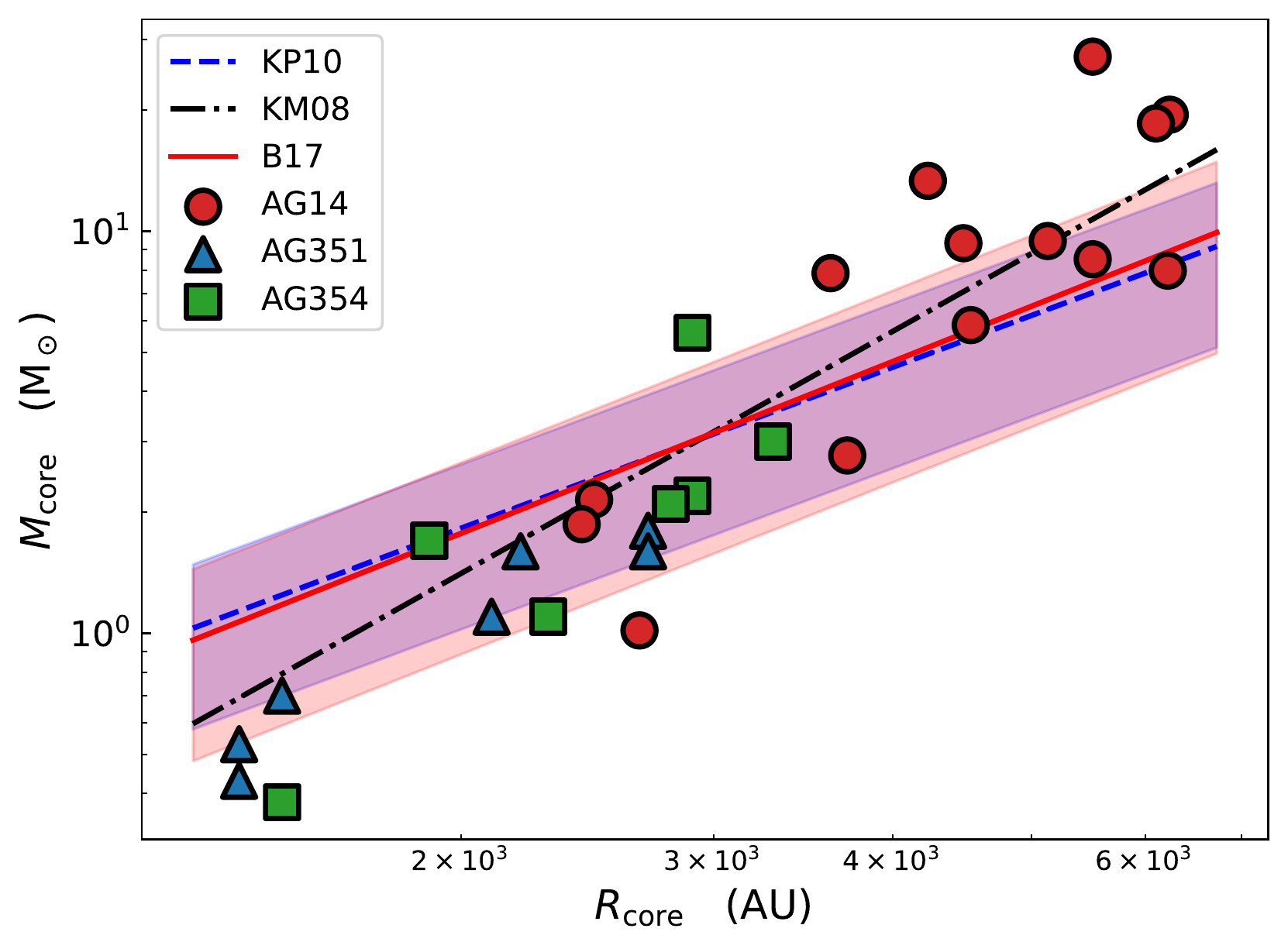}
\caption{ Core masses vs sizes in AG351 (blue triangles), AG354 (green squares), and AG14 (red circles). The thresholds for high-mass star formation described by \cite{Krumholz08}, \cite{Kauffmann10}, and \cite{Baldeschi17} are shown with the dash-dotted black curve, the dashed blue curve, and the solid red line, respectively. For the last two relations we also indicated uncertainties with shaded blue and red areas, which reflect differences in the dust opacity values used here and in those works. The uncertainties on \mcore are $30$\% (AG351 and AG354) and 38\% (AG14). 
\label{MassRadius}}
\end{figure}

In Fig. \ref{MassRadius} we compare the masses and sizes of the identified cores in AG351, AG354, and AG14 ({at} $T_\mathrm{dust} = 10 \, \rm{K}$). Cores in AG14 appear on average larger and more massive than in the other two clumps, as expected by the availability of a larger mass reservoir, since AG14 is a factor of $\approx 30$ more massive than the other two sources. In the figure, we also report several estimations of the threshold for high-mass star formation in the mass-size space. \cite{Krumholz08} derived analytically the surface density limit of $\Sigma \approx 1 \, \rm g \, cm^{-2}$, which roughly translates into $M /\rm M_\odot = 15 \times 10^3  (R/pc) ^{2}$. From observational data of several IRDCs, \cite{Kauffmann10} derived the relation  $M /\rm M_\odot = 870 \, (R/pc) ^{1.33}$, whilst more recently \cite{Baldeschi17} reported $M /\rm M_\odot = 1282 \, (R/pc) ^{1.42}$, based on the analysis of clouds in the Herschel Gould Belt survey. The most massive cores in AG14 sit well above all the relations here studied, and have therefore the potential to form high-mass stars in the future. However, since their masses are $M_\mathrm{core} \approx 10-30 \, \rm{M_\odot}$, they still need to accrete significant mass from the surrounding environment, unless the star formation efficiency is locally high. 
\par
 {As previously noted, the angular resolutions of the observations of the three clumps are well matched to their distinct distances. However, the angular maximum recoverable scale is  approximately the same for all the sources, which means that more large-scale flux is recovered in AG351 and AG354 with respect to AG14. This might affect the comparison between the core masses, which could be overestimated in AG351 and AG354 with respect to AG14. This however would not affect our conclusion that cores in the last clump are more massive than in the first two sources.}

\subsection{The clump-to-core scale kinematics\label{kinematics}}
The centroid velocity map obtained fitting the \olineh data, shown in Fig. \ref{MCweeds_Results}, suggests a complex kinematics of the source, as indicated by the presence of several velocity components at many positions. In order to investigate the kinematics of AG14 at the clump scale, we have used ALMA Band 3 observations of the \nnhp (1-0) transition. {As illustrated in Sect. \ref{Intro}, this transition is better suited to trace the gas at larger scales than the \olineh data. Furthermore, the Band 3 data} have a spatial resolution of $11200 \rm AU \times 6180  AU$  ($0.05 \, \rm pc \times 0.03 \, pc$), and they were acquired including Total Power observations {(which are not available in the Band 7 dataset), which increases the sensitivity to the large scale emission.} These observations are {therefore} ideal to probe the large-scale kinematics of the gas in which the cores identified in \ohhdp are embedded.
\par

\begin{figure*}
\centering
\includegraphics[width=\textwidth]{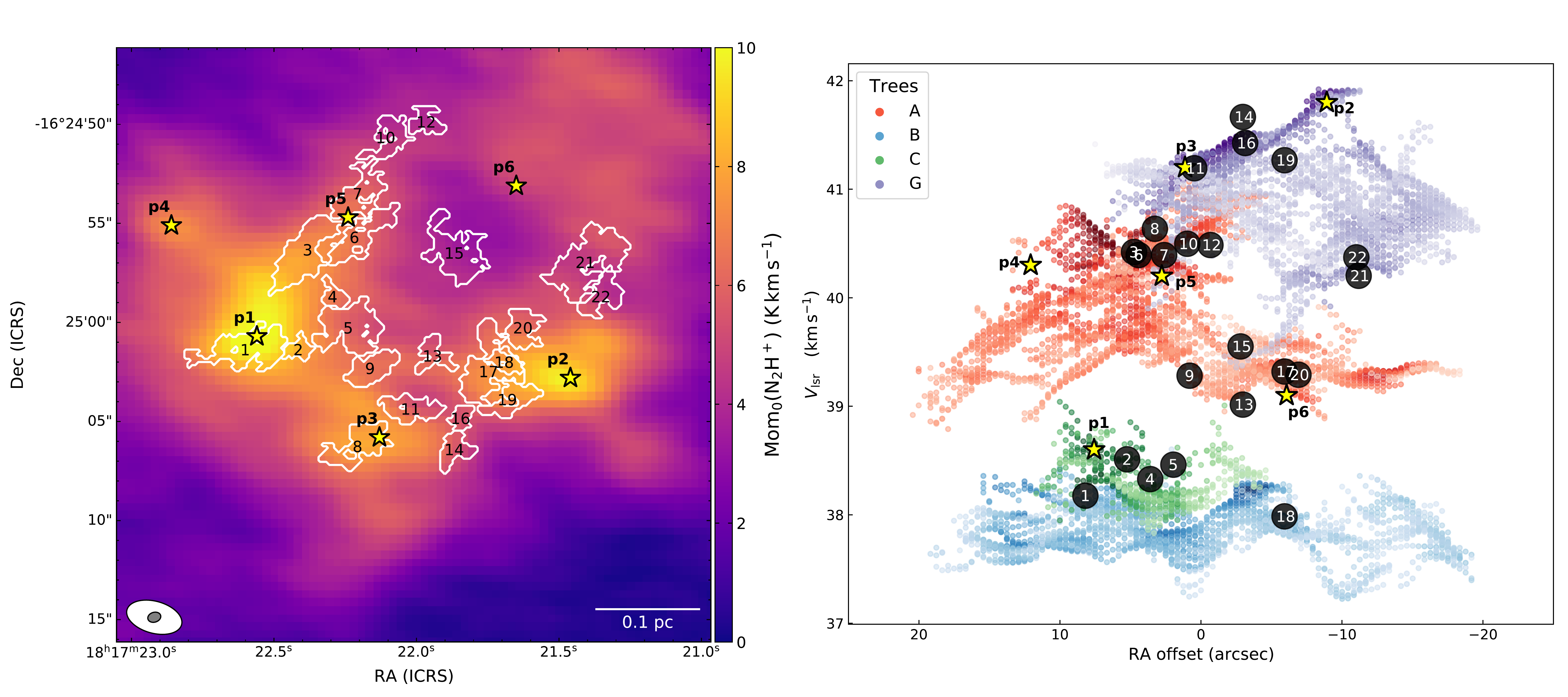}
\caption{\textit{Left panel:} the colorscale shows the integrated intensity of the isolated \nnhp hyperfine component $F_1 = 0 - 1$. The white contours show the cores identified in \ohhdp (labelled with numbers), and the star symbols the positions of the protostars (labelled in bold face). The scalebar is indicated in the bottom-right corner. The white ellipse shows the beam size of the \nnhp data, whilst the smaller grey ellipse represents the \ohhdp beam size. \textit{Right panel:} position-velocity plot from the results of \textsc{acorns}. The plot is the projection in RA-velocity space of the 3D plot shown in Fig. \ref{3d_screenshot}. Right-ascension values are given as offsets with respect to the centre of the FoV (position $\rm RA = 18^h17^m22^s.05$, $\rm Dec = -16\text{\degree} 25' 01''.7$). The four main trees identified by \textsc{acorns} are shown with distinct colours, and the colorscale is determined by the peak intensity of the line, stretching linearly between $0\,K$ and 90\% of the maximum $T_\mathrm{peak}$ in each tree. The positions of the prestellar and the protostellar cores are shown with the black circles and the yellow stars, respectively. Labels correspond to those in the left panel. \label{n2hp_mom0}}
\end{figure*}

We focused on the isolated hyperfine component  $F_1 = 0 - 1$ of the \nnhp (1-0) transition, which is supposed to be optically thin or only moderately optically thick even at the high column densities found in high-mass star-forming regions (see for instance \citealt{Sanhueza12, Barnes18, Fontani21}). The integrated intensity of this component is shown in the left panel of Fig. \ref{n2hp_mom0}, where we overlap also the contours of the \ohhdp cores and the positions of the protostellar objects (star symbols). The field-of-view have been cut to the central $40''\times 40''$, focusing on the map area covered also by the Band 7 data. The \nnhp emission is extended over almost the whole map coverage. The morphology appears filamentary, with several clumpy peaks of emission. Several of these peaks coincide with the position of protostellar candidates.
 \par
 
\begin{figure*}
\centering
\includegraphics[width=\textwidth]{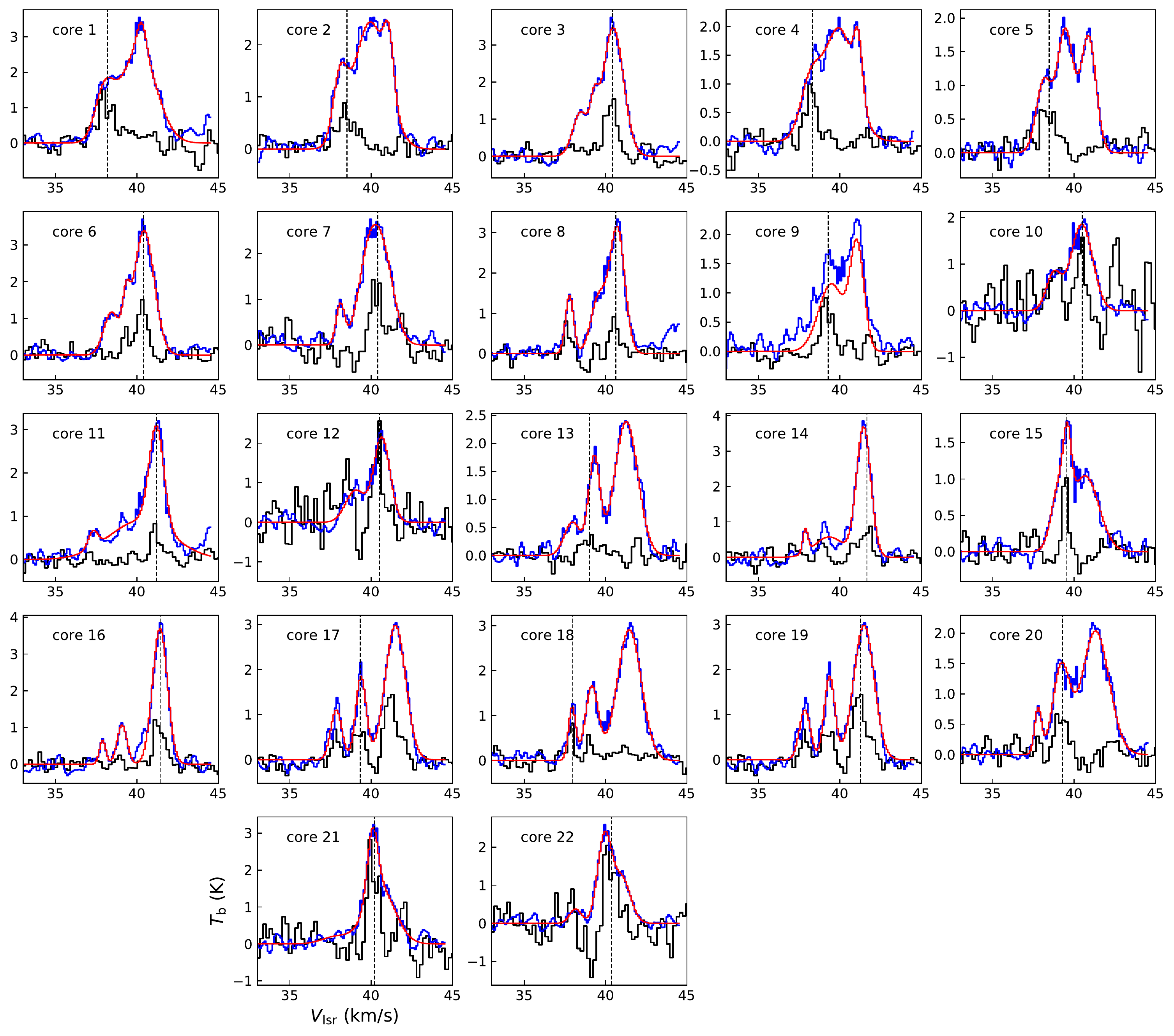}
\caption{Comparison of the \nnhp (1-0), in blue, and the \olineh spectra, in black, at the core positions. The spectra are taken at the position of the peak of \ohhdp integrated intensity. The red curves show the best model obtained fitting the \nnhp data with three Gaussian components. The centroid velocity of the \ohhdp is shown with the vertical dashed line in each panel.  \label{n2hp_oh2dp_spectra}}
\end{figure*}

By a visual inspection of the datacube, it appears that three velocity components are present on a large extension of the source. We have hence proceeded with a three-components Gaussian fit using the \textsc{pyspeckit} package \citep{Ginsburg11}. The technical details of the fitting routine are described in Appendix~\ref{app:fit}. Figure \ref{n2hp_oh2dp_spectra} shows the comparison of the spectra of \ohhdp and \nnhp at the peak of the \ohhdp integrated intensity, for each core identified in Sect. \ref{scimes}. The correspondence between the two species is remarkable. Every velocity components seen in \ohhdp is associated also with a \nnhp component, whilst the opposite is not true. Furthermore, for corresponding components, \olineh tends to present narrower linewidths with respect to the \nnhp line. These findings suggest the following scenario: over the whole clump, at least three gas components (separated in velocity usually by $0.5-1.0\,$\kms) are visible, as traced by \nnhp, an abundant molecule that probes gas densities of $n \gtrsim 10^4 \, \rm cm^{-3}$. Within these large scale structures, cores are formed, with significantly higher densities ($n > 10^5 \, \rm cm^{-3}$, see also Table \ref{CoreProp2}). The gas within the cores is hence cold and dense, and it excites the \olineh emission. Arising from a more quiescent medium, the \olineh spectra are narrower than the \nnhp ones, which instead are associated with larger scale, more turbulent gas, as suggested by the broader linewidths of this species. \par
The \nnhp data allow us to study not only the kinematics within the cores (better traced by the \ohhdp data), but also that of the intraclump gas in which the cores are embedded, since we are able to link kinematically each core with one \nnhp component. In order to investigate the gas structure in ppv space, we used the Agglomerative Clustering for ORganising Nested
Structures (\textsc{acorns}; \citealt{Henshaw19}). Similarly to \textsc{scimes}, \textsc{acorns} is a hierarchical clustering algorithm, which identifies structures and their hierarchical links in position-position-velocity space. However, unlike \textsc{scimes},  \textsc{acorns} is designed to work on already decomposed data. In other words, instead of working on the observed datacubes, it operates on the fitting results of the multi-component Gaussian fit previously described. The technical details about the \textsc{acorns} clustering are given in Appendix~\ref{app:fit}. Using the terminology of \cite{Henshaw19}, \textsc{acorns} finds a forest of 18 trees in total, four of which contains $\approx  70$\% of all data-points, and $\approx 80\%$ of the total flux. These trees present also the most complex hierarchical structures, containing each between two and seven leaves. \par

\begin{figure*}
\begin{interactive}{js}{AG14_n2hp_light.html}
\includegraphics[width=\textwidth]{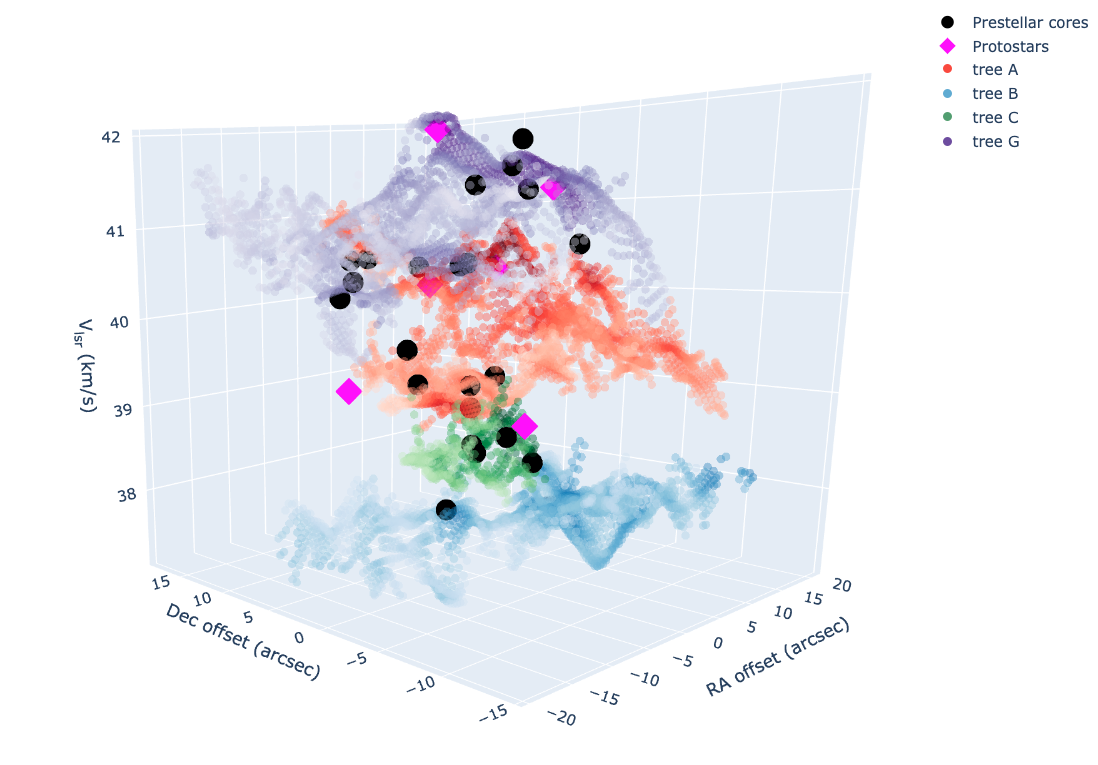}
\end{interactive}
\caption{Screenshot of the ppv distribution of the four main trees identified by \textsc{acorns}, shown with distinct colors (A: red, B: blue, C: green, G: purple). The colorscale within each tree is determined by the peak intensity of the \nnhp component. The black dots show the ppv position of the \hhdp cores (taken at the peak of the line integrated intensity), whilst the magenta diamonds represent the positions of the protostellar cores, following \cite{Li20}. A permanent copy of this interactive plot is available at: \url{http://theory-starformation-group.cl/sbovino/AG14_n2hp_light.html}. {The interactive version allows the reader to rotate the figure, to zoom in and out, and to see the numeric labels of the protostellar and prestellar cores by moving the cursor over.}  \label{3d_screenshot}}
\end{figure*}

In Fig. \ref{3d_screenshot} we show a screenshot of the 3D ppv plot of the four main trees found by \textsc{acorns}, together with the positions in ppv space of the prestellar and protostellar cores. For the prestellar ones, we use the positions of the peak intensity of the \ohhdp integrated intensity within each core, and the centroid velocity at the same position obtained with \textsc{mcweeds} (see Sect. \ref{mcweeds}). The properties of the protostellar cores are derived from \cite{Li20}, who used \dcop, \nndp, or $\rm C^{18}O$ data to infer the systemic velocity values\footnote{A 3D interactive copy of this figure is permanently mantained at: \url{http://theory-starformation-group.cl/sbovino/AG14_n2hp_light.html}.}. A two-dimensional RA-velocity plot of the same data is shown in the right panel of Fig. \ref{n2hp_mom0}.

\par
All the four main trees identified by \textsc{acorns} appear associated with prestellar cores, and at least three of them host protostellar sources. These findings suggest that all of these structures have been or still are active in star formation. The structure at lower \vlsr values (shown in blue in Fig. \ref{3d_screenshot}, labelled B in Fig. \ref{acorns_tot}) is the most coherent in velocity, as it span less than $1.0\,$\kms, despite extending over $\approx 0.75\, \rm pc$ on the plane of the sky. It is also the most quiescent in star formation activity, based on the fact that it is associated with only one prestellar core, and no protostar. On the contrary, the tree coloured in green in Fig. \ref{3d_screenshot} (cluster C in fig. \ref{acorns_tot}), despite having a physical size of $\approx 0.15 \, \rm pc$, contains four cores identified in \ohhdp and one protostellar core. The presence of tracers of both protostellar activity and of cold and dense gas suggest that the star formation is still on-going, and that the protostellar object is very young. The position of the protostellar core p1 is found in fact very close to the \ohhdp core 1, which hints to the fact that the protostellar envelope is still cold and dense enough to have a detectable abundance of \ohhdp. \par

The remaining two trees (labels A and G in Fig. \ref{acorns_tot}; shown respectively in red and  purple in the right panel Fig. \ref{n2hp_mom0} and in Fig. \ref{3d_screenshot}) show a more complex and overlapping structure  in ppv space, and they represent the most dynamically active part of the IRDC clump. At the same time they contain the large majority of cores identified in \ohhdp and three protostellar cores. This is indicative of the fact that this region of the clump is dynamically very active. 
\par
The two protostellar cores p4 and p6 are not associated in ppv space with any of the {four} main {trees} identified in \nnhp. After checking the whole cluster hierarchy found by \textsc{acorns}, however, p4 appears embedded in one of the minor {trees} identified (labelled as 'r' in Fig. \ref{acorns_tot}). The protostellar core p6, instead, has not correspondence in the forest identified by the algorithm. It still emits in the \nnhp (1-0) transition (as can be seen in the integrated intensity map shown in Fig. \ref{n2hp_mom0}), but with low flux, hence not fulfilling the S/N threshold that we require in the fitting algorithm. {A possible interpretation of this observational evidence is that p6} has still an envelope, but this {has} been significantly cleared out by the protostellar activity, suggesting that this could be a more evolved protostar with respect to the others. 
\par
The tree labelled G (shown in purple in the right panel of Fig. \ref{n2hp_mom0} and in Fig. \ref{3d_screenshot}) is one of the largest {identified trees}, as alone it contains more than $20$\% of the total data-points and $\approx 33$\% of the total flux. It also presents {a significant shift in velocity}, extending from $\text{\vlsr} = 39.5 \,$\kms to $42\,$\kms. We now focus on its part connecting the two protostars p2 and p3 (see Fig.\ref{n2hp_mom0}), which presents the brightest peak intensities of the \nnhp line. This section looks like a filament, elongating between the two protostellar cores, and containing four cores identified in the \ohhdp data (n. 11, 14, 16, and 19). The  velocity is increasing from p3 towards p2. 
\begin{figure*}
\centering
\includegraphics[width=\textwidth]{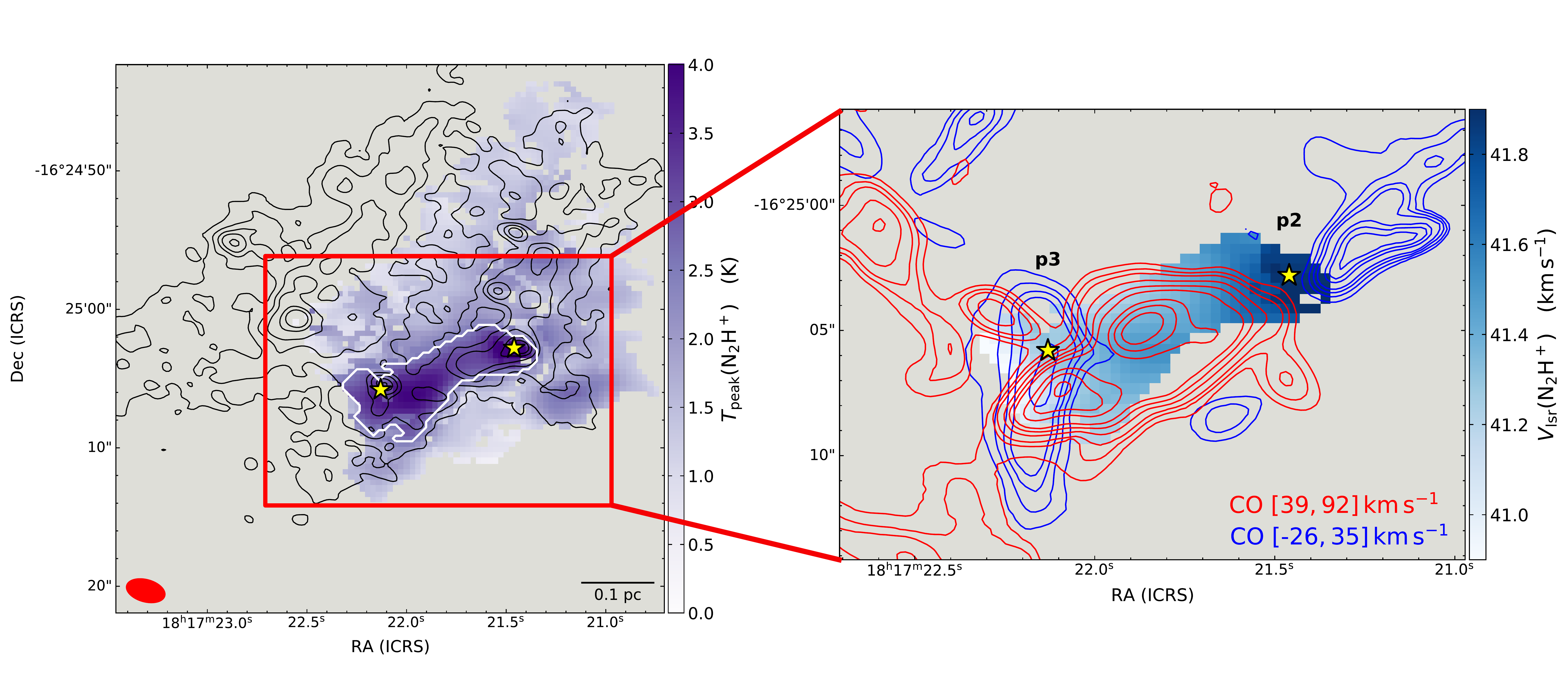}
\caption{\textit{Left panel:} The colorscale shows the peak intensity of the \nnhp components associated with tree G, and the white contour shows the mask (corresponding to $T_\mathrm{peak}> 2.5 \, \rm  K $) used to identify the filament studied in Sect. \ref{kinematics}. The black contours show the continuum emission at 1.34$\,$mm, at levels $[0.5,1.0,1.7,2.5,4.2,5.9]\, \rm mJy/beam$ (from \citealt{Sanhueza19}). The red box indicates the position of the zoomed area shown in the right panel.  \textit{Right panel:} \vlsr map of tree G within the white contour of the left panel, with CO emission overlaid in contours \citep[from][]{Li20}. Contours are computed in the velocity indicated in the bottom-right corner, at levels: for the blue component, $[1.08, 1.62, 2.16, 2.7 , 3.78, 4.32, 4.86]\, \rm Jy\, beam^{-1} \,$\kms; for the red component, $[ 1.33 ,  1.995,  2.66 ,  3.99 ,  5.32 ,  6.65 ,  9.31 , 10.64 ,11.97 ]\, \rm Jy\, beam^{-1} \,$\kms. The positions of protostars p2 and p3 are shown with yellow stars in both panels.\label{tree7}}
\end{figure*}
\par
The left panel of Fig.~\ref{tree7} shows the map of peak intensity of points belonging to tree G. The material surrounding and linking the two protostar emits the brightest lines (with $T_\mathrm{peak}>2.5 \, \rm K$) detected in the source. In the right panel of Fig. \ref{tree7}, we show the \vlsr map of this tree, with overlaid the contours of the outflows detected in the region by \cite{Li20}. The filamentary structure stretching between the two protostellar cores is found in correspondence with the red lobe of the outflow powered by protostar p2. However, the two features cannot coincide spatially, since their velocities are opposite: the outflow is red-shifted, and it has velocities higher than the local standard of rest velocity of protostar p2 ($V_\mathrm{lsr} = 41.8 \,$\kms, according to \citealt{Li20}), whilst the gas traced by the \nnhp line is found at lower (blue-shifted) velocities than that of the protostar. Furthermore, the four \hhdp cores embedded in the gas present low velocity dispersion ($\sigma_\mathrm{V} =0.27-0.34 \,$\kms) and transonic turbulent Mach number ($\mathcal{M}= 1.2 -1.6$, see Table \ref{AveSpectra_par}), suggesting that the dense gas embedded in the filamentary structure is still cold and quiescent, and unperturbed by outflows. \par
{We can thus speculate on the possible scenarios that would give raise to such an observed configuration. A first possibility is simply that we are seeing a bulk motion of the gas. The filament-like structures is moving, and the cores embedded into it participate to this motion. A second possibility is that the gas is flowing towards the protostellar core p2 (i.e. towards increasing velocities), and it accretes material onto the protostar, which in turn powers the bipolar outflow. In any case, the red outflow lobe of p2 and the filament-like structure are found on two distinct planes, which intersect at the position of the protostar, and they appear overlapping in RA-Dec space only due to projection effects.
 }
 \par
{Regardless of the real configuration, the observations unveil a gas flow along the filamentary structure, and it is possible to evaluate the mass-flow rate associated to it. If the correct scenario is the second we proposed, we can interpret this quantity as a mass accretion rate onto the protostar p2. To perform the calculation, we consider the filament} as limited to those positions where $T_\mathrm{peak}>2.5 \, \rm K$, since we want to focus on the denser portion of the gas traced by the \nnhp (1-0) line. This region, shown with the white contour in the left panel of Fig. \ref{tree7}, has a width of $0.08-0.12 \rm \, pc$, which are typical values  for filaments (see e.g. \citealt{Arzoumanian11, Arzoumanian19, Palmeirim13, Sabatini19}), and a length of $\approx 0.26 \rm \, pc$. It spans $1\,$\kms in velocity, from $40.9\,$\kms close to the protostar p3 to   $41.9\,$\kms around p2. The velocity gradient is hence $\nabla V = 3.85 \, \rm km \, s^{-1} \, pc^{-1}$. {T}o estimate the total mass of the filament ($M_\mathrm{fil}$), we employ again Eq. \ref{Mdust}. We use the continuum emission detected in Band 6 at $1.34\, \rm mm$, since its FoV and resolution are closer to the Band 3 data with respect to the Band 7 ones. The flux density contained within the mask shown in Fig. \ref{tree7} is $F_\mathrm{fil} = 85 \rm \, mJy$. A significant contribution to this flux level comes from the bright emission of the cores p2 and p3, which are $16$ and $15\rm \, mJy$, respectively \citep{Sanhueza19}. In these cores, the emission likely arises from the warmer envelope surrounding the protostellar object, and we hence subtract it from $F_\mathrm{fil}$, since we are interested on the flow of gas not associated with the envelope of the protostars. The dust opacity at $1.34\, $mm, computed following Eq. \ref{kappa}, is $0.81 \rm \, cm^{2} \, g^{-1}$. Assuming $T_\mathrm{dust} = 10\, \rm K$, we obtain $M_\mathrm{fil} = 57 \, \rm M_\odot$. The mass accretion rate is then $\dot{M}_\mathrm{acc} = M_\mathrm{fil}  \times \nabla V  = 2.2 \times 10^{-4} \, \rm M_\odot \, yr^{-1}$. {We stress again that this is the rate at which the mass flows along the filamentary structure. The scenario in which it actually corresponds  to an accretion motion is only one of the possibilities that would explain the observations. In order to definitely assess if this is the case, more information, in particular on the protostars (i.e. their masses, luminosities, evolutionary stages,...) would be helpful.}\par
In the following, we discuss the sources of uncertainties that affect {the physical quantities just determined.} First of all, there is the uncertainty on the mass, which accounts for $\approx40$\% (see Sect. \ref{contBand7}), that comes from uncertainties in the dust-to-gas ratio, in the source's distance, and in the dust opacity. Furthermore, the inclination $i$ of the filament with respect to the plane of the sky is unknown, and it affects the value of $\dot{M}_\mathrm{acc}$ by a factor $\tan{i}$ (see e.g. \citealt{Chen19}). If the inclination varies in the range $30-60$\degree, the derived value of the accretion rate changes up to $70$\%. Due to these considerations, with a conservative approach we assume that the derived $\dot{M}_\mathrm{acc}$ value is correct within a factor of two. Within the uncertainties, the value we found is in agreement with measurements in similar sources: for instance, \cite{Lu18} found  $\dot{M}_\mathrm{acc} = (1-2)\times 10^{-4} \, \rm M_\odot \, yr^{-1}$ in filaments belonging to four high-mass star-forming regions, whilst \cite{Chen19} derived $\dot{M}_\mathrm{acc} = (0.2 - 1.3) \times 10^{-4} \, \rm M_\odot \, yr^{-1}$ in several filaments identified in the infrared dark cloud G14.225-0.506. \cite{Sanhueza21} derived $\dot{M}_\mathrm{acc} = (0.9- 2.5) \times 10^{-4} \, \rm M_\odot \, yr^{-1}$ in a hot core embedded in the high-mass star-forming region IRAS 18089-1732, even though at smaller spatial scales ($\approx 10000\,$AU). In \cite{Li22}, authors studied the accretion in a filament in the high-mass star-forming region NGC6334S, deriving $\dot{M}_\mathrm{acc} = 0.3\times 10^{-4} \, \rm M_\odot \, yr^{-1}$. {Furthermore, this value is also in agreement with the results of numerical simulations \citep[see e.g.][]{Wang10, Kuiper16}.}\par
The critical line mass of a filament, in the approximation of isothermal cylindrical shape, is \citep{Ostriker64}:
\begin{equation}
m_\mathrm{c} = \frac{2 c_\mathrm{s}^2}{G} = 17 \left ( \frac{T_\mathrm{K}}{10 \rm \, K} \right ) \rm \, M_\odot \, pc^{-1} \; .
\end{equation}
The line mass of the filamentary-like structure in AG14 is $m= M_\mathrm{fil}/L_\mathrm{fil} = 220\rm \, M_\odot \, pc^{-1}$, i.e. significantly higher than its critical value, which suggests that this structure is out of hydrostatic equilibrium. {One could naturally wonder whether this is consistent with the possible scenario of accretion flow that has been discussed. In particular, it is worth comparing the timescales for accretion ($t_\mathrm{acc}$) and free-fall collapse ($t_\mathrm{ff}$), at least in terms of orders of magnitude. The former can be approximated by the ratio between the mass reservoir and the accretion rate: $t_\mathrm{acc} = M_\mathrm{fil}/\dot{M}_\mathrm{acc}= 2.6 \times 10^5 \, \rm yr$. To estimate the time necessary for a filament to fully collapse onto its axis, we use Eq. 18 of \cite{Hacar22}, which in turn was derived from \cite{Pon12,Toala12}:}
\begin{equation}
t_\mathrm{ff} = 1.9 \left (\frac{ L_\mathrm{fil}}{FWHM_\mathrm{fil}}\right )^{0.5} \left ( \frac{n_0}{10^3 \, \mathrm{cm^{-3}}}\right ) ^{-0.5}\, \rm Myr \; ,  
\end{equation}
{where $L_\mathrm{fil}/FWHM_\mathrm{fil} = 2.6 $ is the filament aspect ratio (i.e. the ratio between its length $L_\mathrm{fil} = 0.26\,$pc and its width $FWHM_\mathrm{fil}=0.1\, $pc), and $n_0$ is the filament central density at the spine. Since we are interested only in a rough estimation of this quantity, we test the range of densities found in the cores\footnote{{This values are also consistent with the average density of the filament, computed assuming that it is a perfect cylinder of length $L_\mathrm{fil}$ and radius $FWHM_\mathrm{fil}$.}}, i.e. $n_0 = 10^5-10^6 \, \rm cm^{-3}$, obtaining $t_\mathrm{ff} = 1-3 \times 10^5 \, \rm yr$. We conclude that the two timescales are comparable, and hence the filament would have time to accrete a significant fraction of its mass onto p2 before collapsing. }


\subsection{Comparison between \ohhdp and \nndp \label{h2dp_n2dp}}

\begin{figure}[!b]
\centering
\includegraphics[width=.5\textwidth]{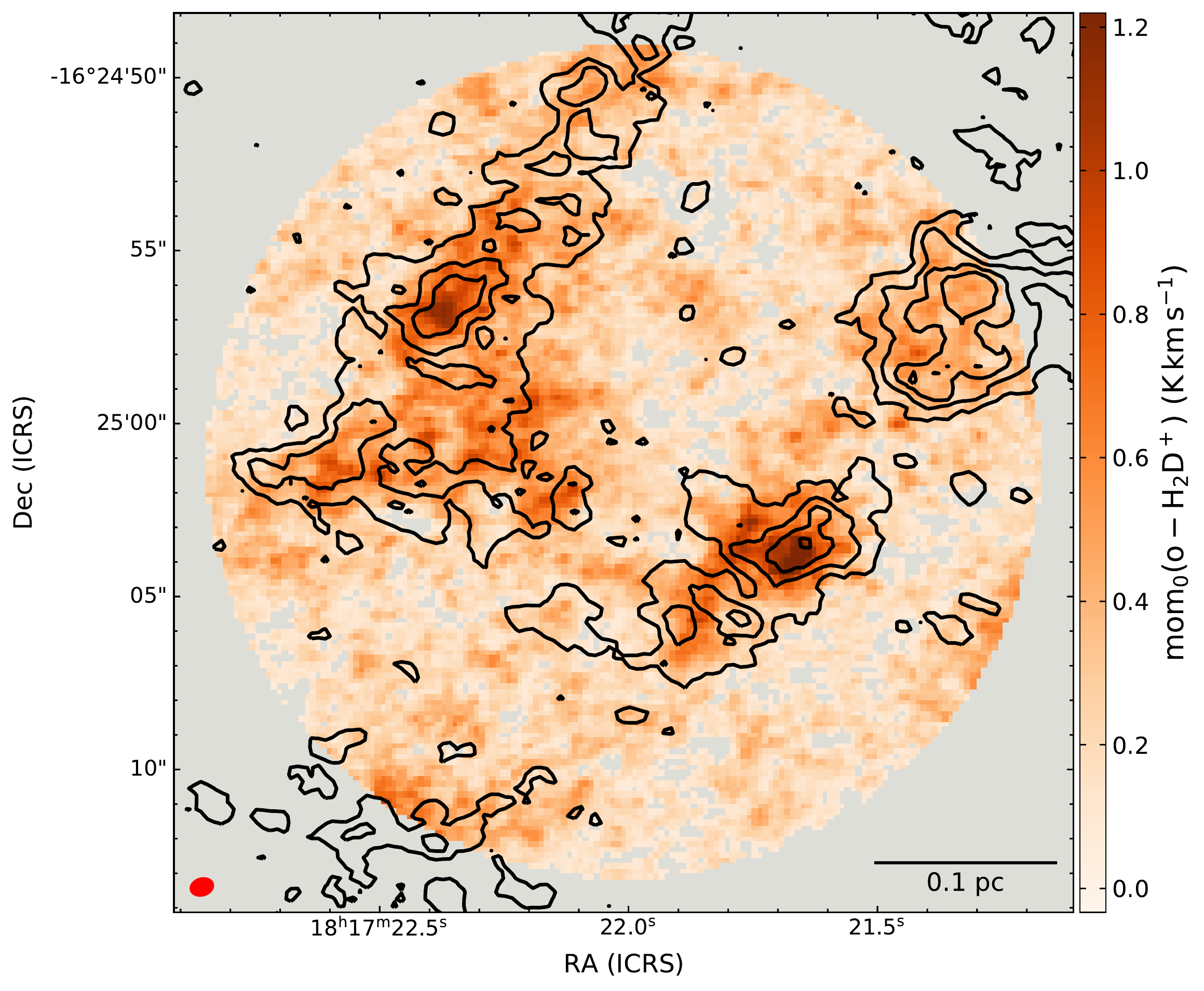}
\caption{The colorscale shows the integrated intensity of the \olineh transition as in Fig.~\ref{Band7_Data}. The black contours shows the  \nndp (3-2) integrated intensity (computed as for \ohhdp, see Sect. \ref{Observations}). The levels are: $[7,11,15,19]\sigma$, where $1\sigma = 100 \rm \, mK \, km \, s^{-1}$. \label{n2dp_mom0}}
\end{figure}
\begin{figure*}
\centering
\includegraphics[width=.9\textwidth]{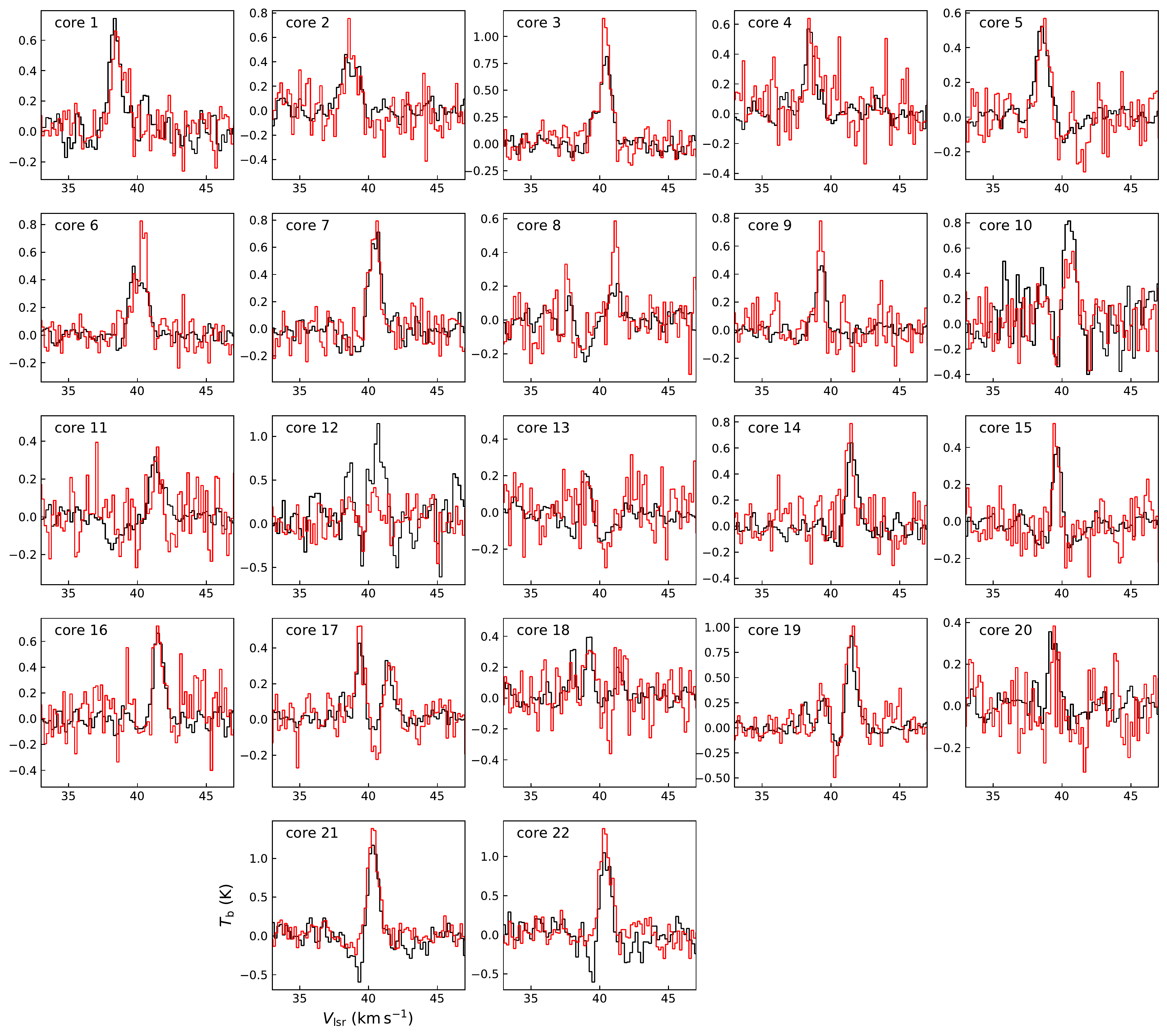}
\caption{Comparison between the average \nndp (3-2) and \olineh lines in each core (labelled at the top-left corner of each panel). \nndp data are shown with the red histograms, whilst \ohhdp ones are shown with the black curves. Note that the \ohhdp data have been smoothed to the same beam size of the \nndp cube before extracting the average spectra, which thus are not identical to the ones presented in Fig. \ref{AveSpectra_h2dp}.
\label{Spectra_h2dp_n2dp}}
\end{figure*}
The cores identified in Sect. \ref{scimes} using \ohhdp data are formed by cold and dense gas that should be in a prestellar stage, even though we have evidence that a minority of cores are found in close proximity to protostellar cores (in ppv space), such as core 1 (close to protostar p1), or cores 3, 6, and 7 (close to p5). We can speculate that in these cases the \ohhdp emission is tracing the part of the protostellar envelopes that is still cold enough that the desorption of CO from the dust grains has not happened yet. This is confirmed by depletion maps derived from $\rm C^{18}O$ (2-1) observations {of AG14} at $1''.3$ of resolution, which show that the depletion factor is high ($f_\mathrm{D}>50$) even around protostellar cores \citep{Sabatini22}. This suggests that ALMA observations at resolution of $\approx 1''$ are tracing the dense and still cold envelope around protostellar objects, where the feedback of the protostar has not affected the gas yet.  However, even the remaining cores could still belong to distinct evolutionary stages. To these regards, {we have mentioned that} \cite{Giannetti19} studied the correlation between the \olineh and the \nndp (3-2) transitions in three clumps embedded in the G351.77-0.51 complex, using single-dish data from APEX. The main result of those authors was an anticorrelation between the abundances of the two molecular species, {possibly due to evolutionary effects. Their} findings hinted to the possibility of using the abundance ratio between \nndp and \ohhdp as an evolutionary indicator. \par
{In Fig. \ref{n2dp_mom0} we show the comparison between the integrated intensities of the \nndp (3-2) and the \olineh transitions. The two tracers appear quite correlated spatially, but some differences are visible. For instance, in the north-west part of the source, the \nndp line has a bright peak, which is not seen in \ohhdp. Furthermore, the \nndp transition seems more extended, even though we must highlight a possible observational bias: despite we excluded the Total Power observations from the Band 6 dataset, its maximum recoverable scale is still almost twice that of the \ohhdp data, hence making the former more sensitive to large-scale emission.}\par
\begin{figure*}
\centering
\includegraphics[width=.9\textwidth]{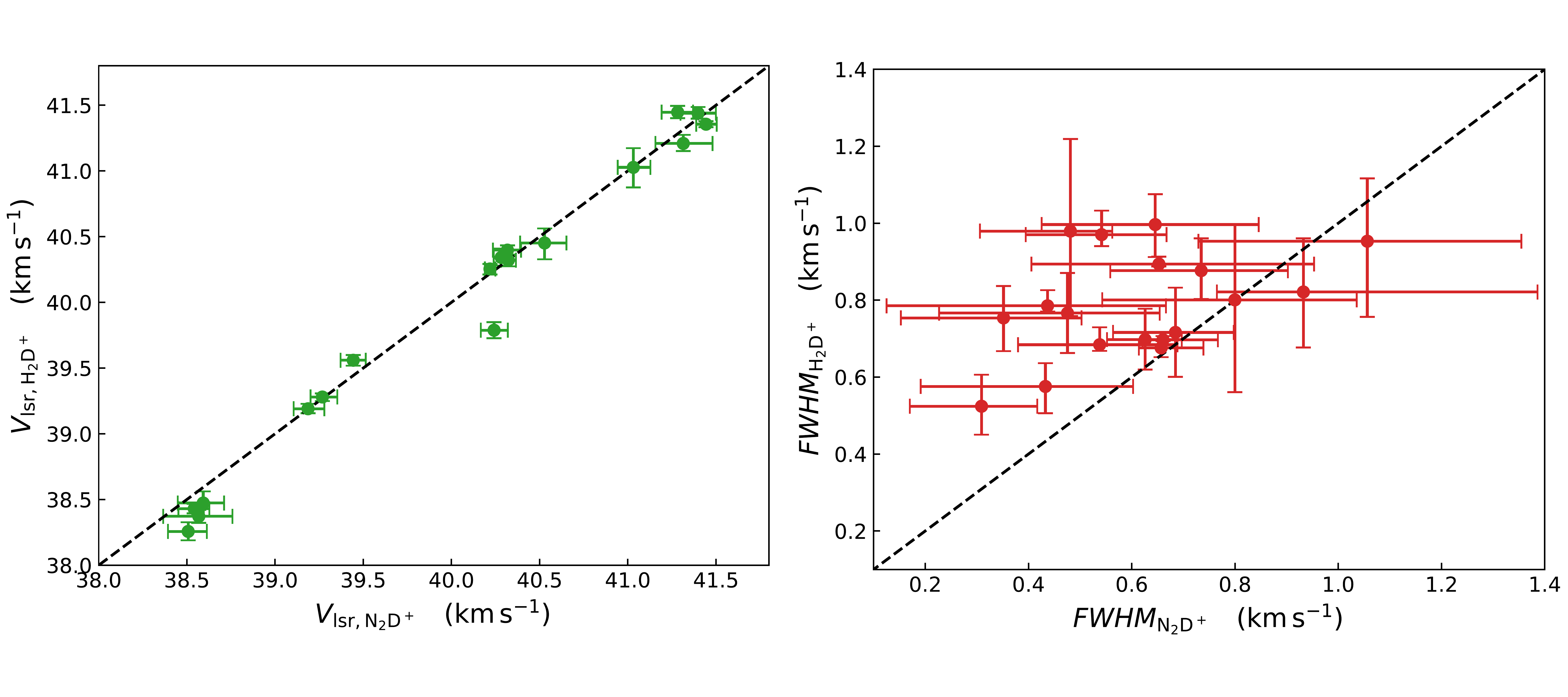}
\caption{\textit{Left panel:}Comparison of the centroid velocities obtained with \textsc{mcweeds} on the \nndp and \ohhdp average spectra in the cores where both species have been detected above the $3\sigma$ level. The black dashed line shows the $1:1$ relation. \textit{Right panel:} Same as in the left panel, but for the linewidths.
\label{h2dp_n2dp_fig}}
\end{figure*}

{In order to further explore the comparison at the core level,} we compared the average spectra of \olineh and \nndp (3-2) in each core. Since the Band 6 data have a lower resolution of $1''.4 \times 1''.0$, we first smoothed the Band 7 data to this beam size, to allow for a proper comparison. {Both spectral cubes have been regridded to the same coordinate grid.} The spectral resolution of the two datasets is comparable ($0.20\,$\kms for \ohhdp and $0.17\,$\kms for \nndp). The comparison of the average spectra is shown in Fig. \ref{Spectra_h2dp_n2dp}. The similarities between the line profiles of the two tracers are remarkable, both in terms of intensity and of line shapes. In four cores (12, 13, 18, and 20) the \nndp transition is not detected above the $3\sigma$ level, but we highlight that the $rms$ of the \nndp spectra is on average $\approx 2.5$ times worse than in the corresponding \ohhdp spectra. \par

We have fitted the average spectra shown in Fig.\ref{Spectra_h2dp_n2dp} with \textsc{mcweeds}. We assumed $T_\mathrm{ex} = 10 \, \rm K$ also for \nndp, for which we take into consideration the hyperfine splitting due to the $^{14}\rm N$ nuclei, according to the CDMS database. Figure \ref{h2dp_n2dp_fig} shows the comparison of the best-fit values for the centroid velocity and $FWHM$ for the two tracers in the cores where both are detected above $3\sigma$. The \vlsr values align very well, considering the uncertainties, with the 1:1 relation (shown with the black-dashed curve), highlighting that the two molecular emissions arise from similar spatial regions within the source. Concerning the linewidhts, the right panel of Fig. \ref{h2dp_n2dp_fig} shows that for 80\% of the cores the \ohhdp transition presents broader lines with respect to \nndp. This can be {partially due to opacity effect, since the \olineh can be moderately optically thick, leading to a 15\% overestimation of the linewidth (see Appendix \ref{missingFlux})}. The presence of the hyperfine splitting in the \nndp (3-2) transition, instead, reduces this problem. Furthermore, we highlight the difference in the critical density of these two tracers (one order of magnitude higher for the \nndp line than for the \ohhdp transition).

\begin{figure}[!b]
\centering
\includegraphics[width=.5\textwidth]{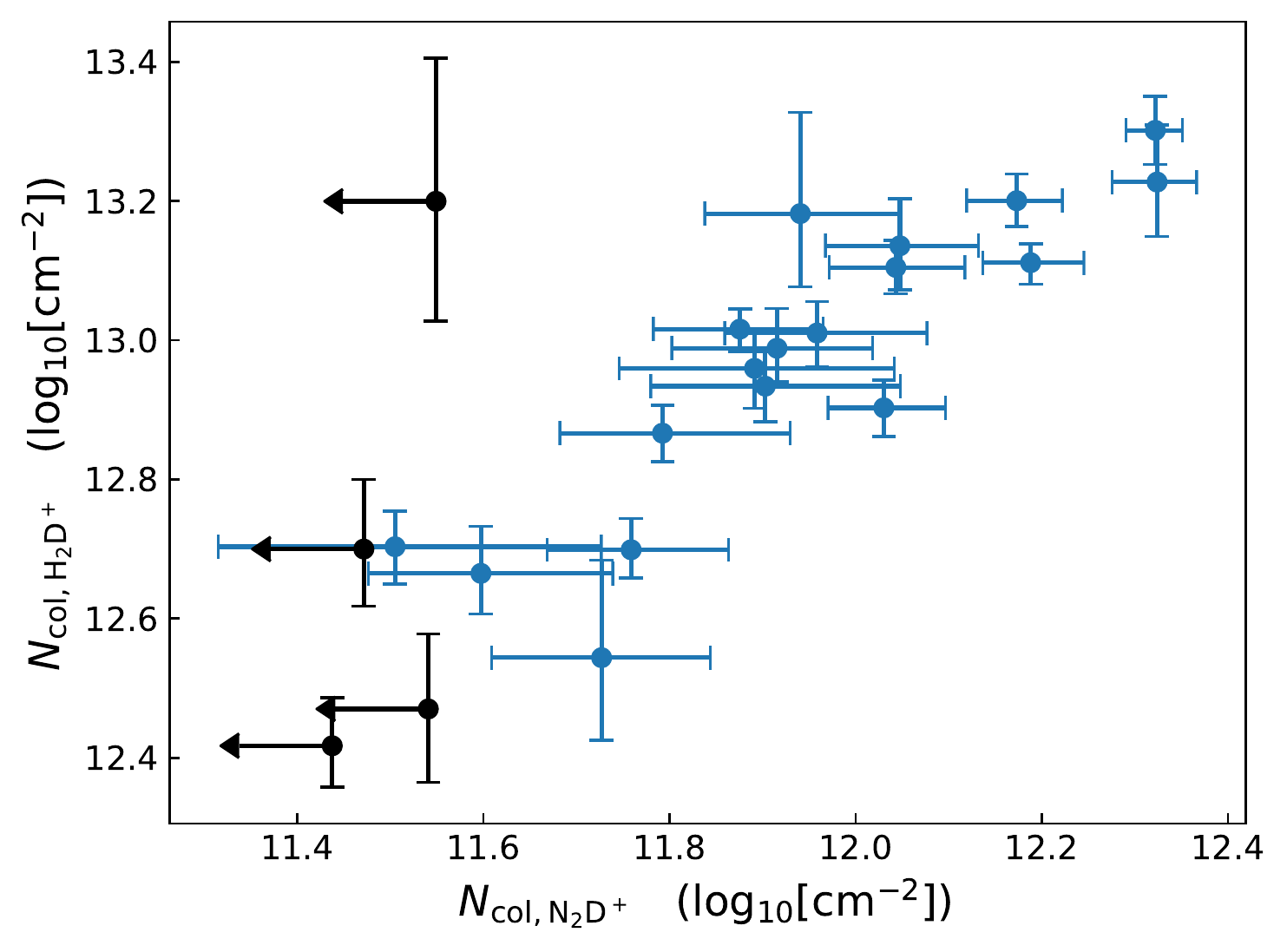}
\caption{Same as Fig. \ref{h2dp_n2dp_fig}, but for the column densities. For the four cores where \nndp is not detected, we show upper limits (in black).
\label{Ncol_h2dp_n2dp}}
\end{figure}

\par
In Fig. \ref{Ncol_h2dp_n2dp} we report the correlation between the column density values of \ohhdp and \nndp. For the cores undetected in \nndp, we report $3\sigma$-upper limits computed based on the $rms$ in each core and the average linewidths of the detected cores ($<FWHM> = 0.62\,$\kms). We found a discrepancy with the anti-correlation trend found by \cite{Giannetti19}, since the column density of \nndp and \ohhdp appear well correlated, for the cores detected in both tracers. However, we highlight a fundamental difference between our analysis and that of \cite{Giannetti19}. Those authors selected cores in continuum emission, and then analysed the molecular emission. Our analysis, instead, is intrinsically biased towards core with bright \ohhdp emission, since we used this species to identify core-like substructures. Furthermore, \cite{Giannetti19} investigated the clump level scales, and therefore the anti-correlation reflected averaged clump properties. In this work, on the contrary, we resolve the core scales in a highly dynamically active environment, which further complicate a direct comparison between the two works.

\par
It is worth commenting on the four cores undetected in \nndp emission. They present narrow \ohhdp lines ($\sigma_\mathrm{V}= 0.18-0.27\, $\kms, smaller than the average velocity dispersion of all cores in the clump), hinting to cold and quiescent gas. According to the analysis of the continuum emission (see Sect. \ref{contBand7} and Table \ref{CoreProp2}), three of them are undetected in continuum emission, and the last one (18) is the least massive of the sample ($M_\mathrm{core} \lesssim 1.0 \rm \, M_\odot$). The non-detection of \nndp can be then explained by two scenarios: \textit{i)} these cores are not dense enough to excite the \nndp transition that has an higher critical density with respect to that of the \olineh line; \textit{ii)} alternatively, the lack of continuum emission can be explained if the gas and dust temperatures are so low ($<10\, \rm K$) that the dust thermal emission at $0.8 \rm \, mm$ is not bright; in this case, the cores would be in early evolutionary stage, and perhaps the \nndp, a late-type species, has not yet formed in detectable quantities. In this case, the core masses estimated in Sect. \ref{contBand7} could be underestimated.

\section{Discussion and Conclusions\label{Conclusion}}
In this work, we have investigated the dynamical and kinematic properties of AG14, from the core to the clump scales, analysing ALMA data at spatial resolution from $\sim$ 2000 to $\sim$ 12000 AU ($0.01-0.06\, \rm pc$). Using Band 7 \ohhdp data, we have identified 22 cores with dendrogram analysis. Comparing their distribution with the dust thermal emission in the same band, most bright continuum peaks are found outside or right at the edge of the \hhdp cores. Several of these peaks are known to be associated with outflow activity, and therefore are likely protostars. The fact that they lack \ohhdp emission can be explained if they are already quite evolved, and the protostellar feedback has heated the surrounding gas above the CO desorption temperature. If CO is back into the gas phase, its fast reaction with \hhdp would lower the abundance of the latter below the detection level. Alternatively, if they are in earlier evolutionary stage and they are still dense and cold, \hhdp could be efficiently transformed into its doubly and triply deuterated forms, or it could deplete due to the depletion of HD itself \citep{Sipila13}. \par
{The identified cores have typical masses of $M_\mathrm{core} \lesssim 30 \, \rm M_\odot$, and they appear subvirial at $T =10\, \rm K$, even though the virial parameters might be underestimated (see Sect. \ref{contBand7} for more details). Our data seem to exclude the existence of HMPCs in AG14, even though our mass values could be underestimated due either to filter-out of large-scale emission by the interferometer, or due to overestimation of the dust temperature. However, the \ohhdp line adds no support for temperatures lower than $10\, \rm K$, unlike in AG351 and AG354, where a significant fraction of pixels presented lines narrower than the thermal broadening at $10\, \rm K$ (see \citealt{Redaelli21}). }
\par
{The \olineh transition at the clump level span a range of $\approx 4\,$\kms in \vlsr, and its morphology suggests that multiple velocity components are present in the source. In order to study the large-scale clump kinematics of the gas in which the identified cores are embedded, we used ALMA Band 3 observations of the \nnhp (1-0), which is} an ideal probe for the large scale kinematics. From the spectral comparison of the two tracers (the first ever done in literature, to our knowledge), we can link kinematically each \hhdp core with one velocity component of the \nnhp spectra. The high density cores are hence formed in the large-scale gas traced by \nnhp, and they inherit its kinematics. The \nnhp lines are on average broader than the corresponding \olineh components, suggesting that the denser gas is more quiescent, as expected from turbulence dissipation. \par
\begin{figure*}[!t]
\centering
\includegraphics[width=.8\textwidth]{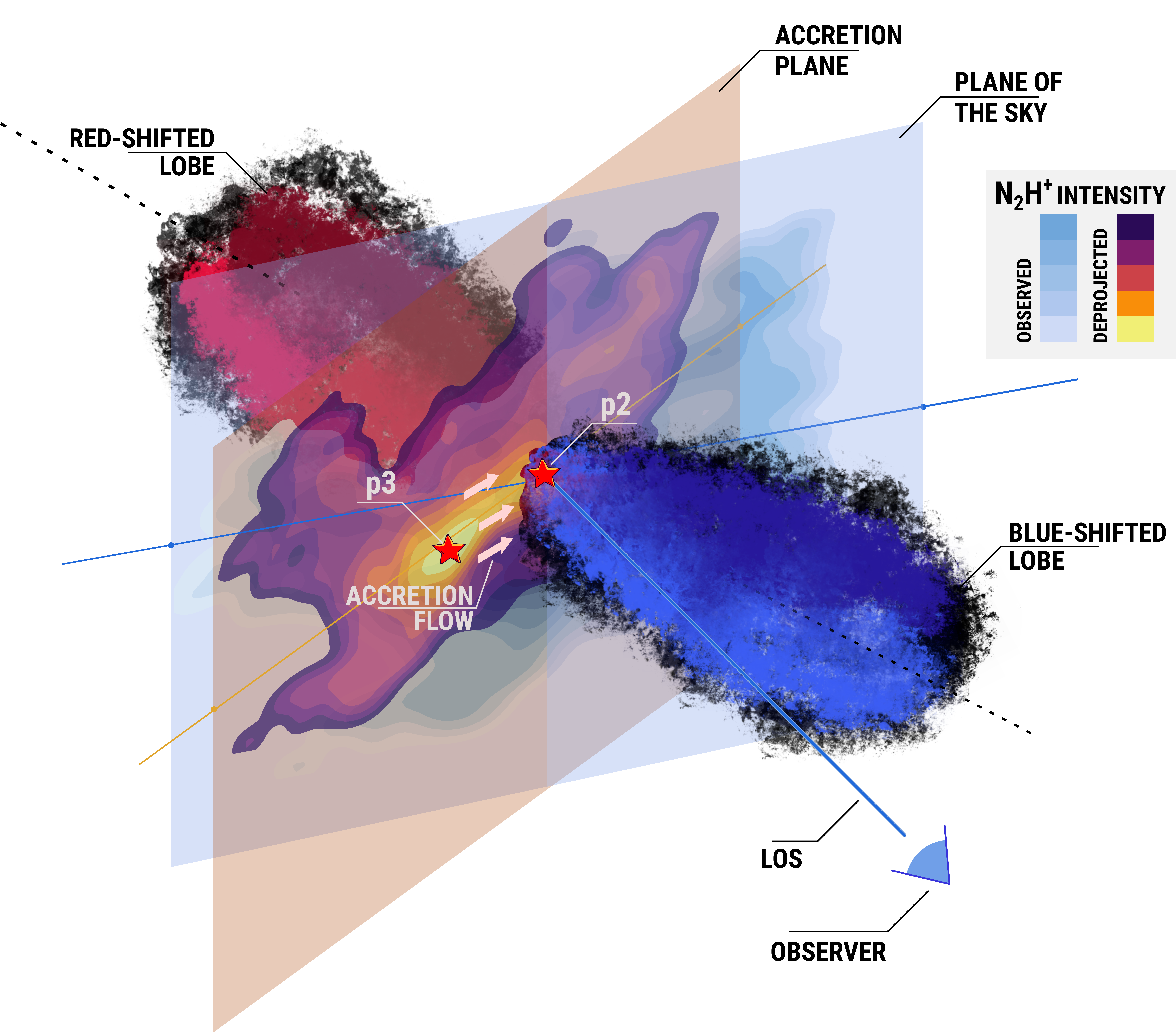}
\caption{ A graphic illustration of the possible 3D configuration that gives raise to the observed structure in Fig. \ref{tree7}. A filamentary structure seen in \nnhp (1-0) emission accretes material towards the protostellar core p2. This launches a bipolar outflow likely almost perpendicularly to the accretion flow. The red lobe, when projected on the plane of the sky, appears to coincide with the filament. \label{Tommaso}}
\end{figure*}
To disentangle the complex kinematics shown by the \nnhp data and to identify its hierarchical structure in ppv space, we have first fitted the isolated hyperfine component $F_1 = 1-0$ using a three-component Gaussian fit, and then we have used the results as input for the \textsc{acorns} package \citep{Henshaw19}. {The four main trees found by \textsc{acorns} are associated with cores identified in \ohhdp emission, and at least three host also protostellar cores, suggesting that all of them are active in star formation.}  One of the trees (labelled B) presents a small velocity gradient ($\approx 1\,$\kms over $0.75\, \rm pc$) and it appears more quiescent than the others, since it contains only one prestellar core. Interestingly, this core (18) is one of those not detected in \nndp, which can be explained if it is at an early evolutionary stage, when \nndp ---a late type molecule--- did not have the time yet to form. This tree can then represent a less evolved component with respect to the others in the clump.\par
The trees labelled A and G are associated with more than 70\% of the \ohhdp cores and three protostellar cores, and they are overlapping and intertwined in ppv space. Such a morphology could be indicative of a sort of competitive accretion scenario, where in the crowded environment of this high-mass clump, multiple low-mass cores ($M_\mathrm{core} < 30 \, \rm M_\odot$) have formed. The intraclump gas in which the cores are embedded could provide the cores with the mass reservoir needed to later form high-mass stars. This is also consistent with the fact that at $10 \, \rm K$ all the cores are subvirial. \par

{The brightest part of tree G, is structured as a} filamentary structure connecting the two protostellar cores p3 and p2. The \nnhp centroid velocity increases from $\sim 41\,$\kms close to p3 to $\sim 42\,$\kms close to p2. On the plane of the sky, this structure overlaps with the red lobe of the CO outflow identified by \cite{Li20}. However, the outflow velocities are opposite to that of the \nnhp filament, since they are redshifted with respect to the systemic velocity of p2 ($42.8\, $\kms). {We have speculated on the possibilities that would explain the observed configuration, and we show one of them} in Fig. \ref{Tommaso}. The filamentary structure seen in \nnhp emission {might be accreting} mass onto the protostellar core p2, which then powers a bipolar outflow in a direction likely perpendicular to that of the accretion flow; the red lobe of the bipolar outflow, when seen projected on the plane of the sky, appears overlapped to the \nnhp feature, but the two are actually separated in 3D space. {Assuming this scenario, we have computed the mass accretion rate along the filamentary structure, obtaining $\dot{M}_\mathrm{acc} = 2.2 \times 10^{-4} \, \rm M_\odot \, yr^{-1}$, expected to be accurate within a factor of two, in good agreement with other observations in similar sources. }From the outflow parameters, \cite{Li20} estimated a mass accretion rate on the protostar p2 of $3-4\times 10^{-6} \rm \, M_\odot \, yr^{-1}$, i.e. approximately two order of magnitude lower than our estimate, but this value depends on several assumption (for instance on the wind velocity and on the ratio between the mass accretion rate and the mass ejection rate). Furthermore, the value of \cite{Li20} represents the accretion rate onto the protostar, whilst we compute the rate onto the core.
\par
In this work, we have shown how ALMA observations of several molecular tracers are a powerful diagnostic tool to investigate the fragmentation and kinematic properties of the high-mass clump AG14. In particular, \ohhdp appears an ideal tracer of the cold and dense gas, and as such it can be used to identify cores likely in an early evolutionary stage. On the other hand, Band 3 data of \nnhp can be used to trace the gas kinematics at clump scales and at the clump-to-core transition, providing important information on the dynamics and accretion properties of the gas from which the cores formed. 

\restartappendixnumbering

   \begin{acknowledgements}
 {The authors thank the anonymous referee for the comments which helped to improve the quality of the manuscript.}  The authors acknowledge Tommaso Grassi for the help in producing Fig. \ref{Tommaso}. ER acknowledges the support from the Minerva Fast Track Program of the Max Planck Society. ER and PC acknowledge the support of the Max Planck Society. SB is financially supported by ANID Fondecyt Regular (project \#1220033), and the ANID BASAL projects ACE210002 and FB210003. PS was partially supported by a Grant-in-Aid for Scientific Research (KAKENHI Number 18H01259 and 22H01271). This research made use of \textsc{astrodendro}, a Python package to compute dendrograms of Astronomical data (\url{http://www.dendrograms.org/})
  \end{acknowledgements}

\software{\textsc{scimes} \citep{Colombo15}, \textsc{mcweeds} \citep{Giannetti17}, PyMC \citep{Patil10}, \textsc{pyspeckit} \citep{Ginsburg11}, \textsc{acorns} \citep{Henshaw19}, \textsc{astrodendro} (\url{http://www.dendrograms.org/})}

\appendix
\section{Cores identification in continuum emission \label{ContCores}}
In Sect. \ref{scimes} it has been discussed how the morphology of the continuum emission and of the \ohhdp integrated intensity do not correlate. To strengthen this point, we have performed a core identification also in the dust thermal emission, similarly to what done in Appendix B of \cite{Redaelli21}. We highlight that \cite{Sanhueza19} already performed a core-finding analysis in the clump, using the continuum data in Band 6, which have a sensitivity $\approx 4$ times higher than the Band 7 data. We however prefer to use the continuum at 0.8$\,$mm to perform the comparison with the \ohhdp analysis, since these two datasets were observed with the same ALMA configuration. 
\begin{figure}[!h]
\centering
\includegraphics[width=.7\textwidth]{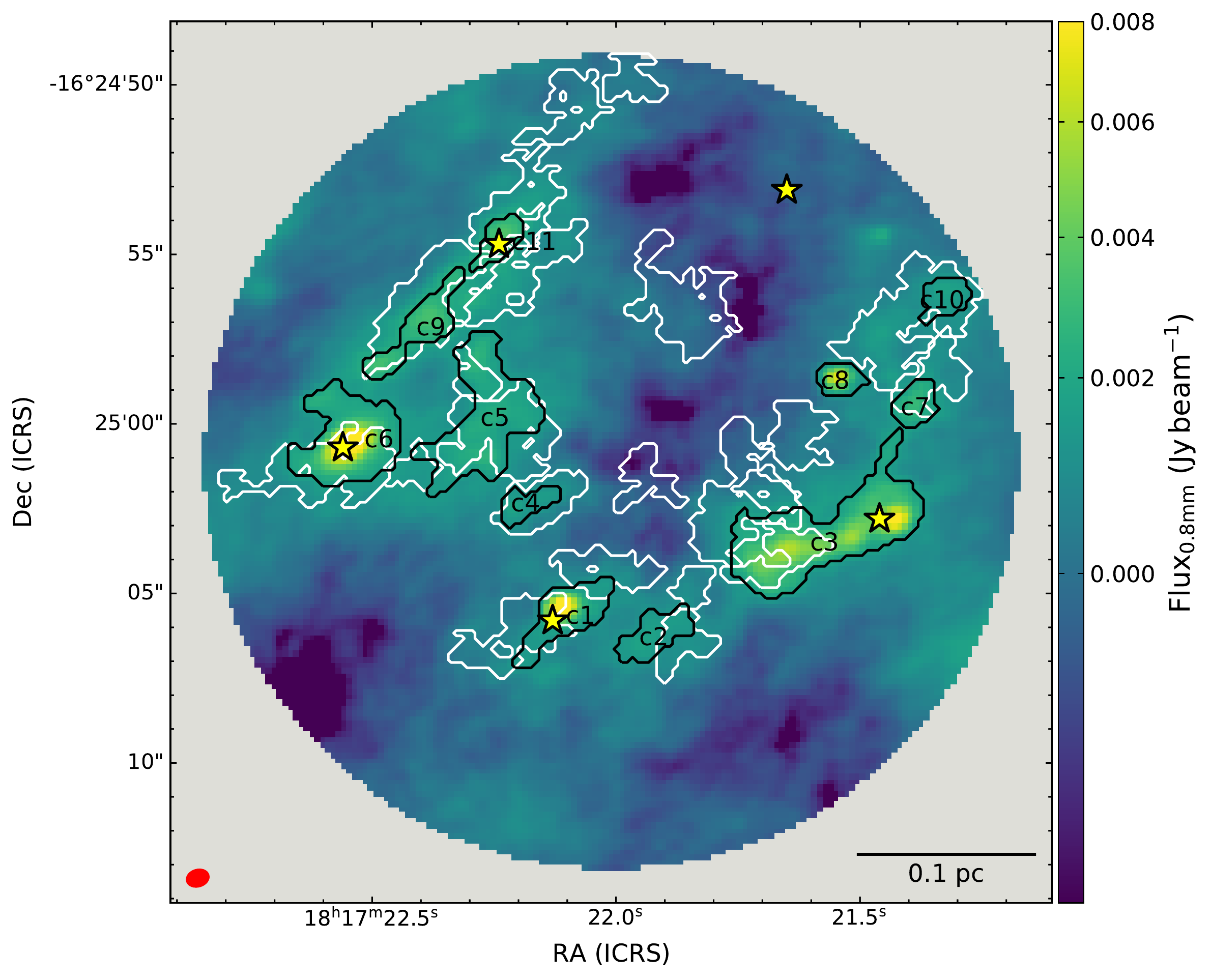}
\caption{The colorscale show the continuum emission in Band 7. The white contours represent the \ohhdp cores, whilst the black ones show the cores identified in continuum, which are labelled based on increasing declination. The beam size and scalebar are shown in the bottom left and right corners, respectively. Cores associated with outflow emission according to \cite{Li20} are shown with stars. \label{fig:contcores}}
\end{figure}
\par
Since \textsc{scimes} works in ppv space, we used the \textsc{python} package \textsc{astrodendro}, on which \textsc{scimes} is based, to analyse the 2D continuum map. Concerning the input parameters necessary to perform the clustering, we set $\Delta_\mathrm{min} = 1 \times rms$ ($rms= 0.5 \, \rm mJy\, pix^{-1}$ for the non primary-beam corrected map); the minimum value to identify structures is $min_\mathrm{val} = 2.5 \times rms$; the identified cores must be larger that three times the beam size, in order to be consistent with the identification of the \ohhdp cores.  \par
With these inputs, \textsc{astrodendro} identifies 11 cores, shown in Fig. \ref{fig:contcores}. Four of them (c1, c3, c6, and c11) are found in correspondence with the protostellar candidates. Five cores (22\%) seen in \ohhdp do not correspond to continuum-identified structures, which suggests that they are in a very early stage, and  their low temperatures translates in low continuum fluxes at $0.8 \, \rm mm$. In turn, continuum core c8 does not overlap with any structure seen in \olineh. This structure corresponds to core 5 in the analysis of \cite{Sanhueza19} and \cite{Li20}. The latter paper does not consider it as associated with clear outflow emission, even though it shows evidence of CO emission at high velocities. Furthermore, \cite{Sanhueza19} lists it among the cores with emission from high-energy transitions. It is hence possible that core c8 hosts a young protostar, that either does not power outflow, or that cannot be detected due for instance to projection effects. \par

Core c3 is the largest one, but it contains three separated flux peaks. It is likely that the algorithm is not able to separate them due to the limited sensitivity of our data. In fact, in \cite{Sanhueza19}  two separated cores were identified in this area in the 1.34$\,$mm continuum emission. In this scenario, core c3 is hence divided in two parts, one which is in a protostellar stage and does not show significant \ohhdp emission; the other  instead is in an earlier evolutionary stage, and it overlaps with several cores seen in \ohhdp.  Core c6 is peculiar, in the sense that it overlaps significantly ($>50$\%) with \ohhdp core 1, and it also contains a protostar. As already suggested in Sect. \ref{kinematics}, these features suggest that this protostar is young, still embedded in a thick envelope that is relatively cold to have a detectable abundance of \ohhdp. \par
In conclusion, more than half of the \ohhdp identified cores overlap with continuum cores by less than 30\% of their physical extension. This is likely due to different evolutionary stages traced by the two dataset. Whilst the \ohhdp emission trace cold gas still relatively undisturbed by protostellar activity, the continuum data cannot distinguish between cores in prestellar and protostellar phase. 
{We note that \cite{Sanhueza19} already identified cores in continuum. We prefer to re-do this analysis, since the Band 6 data used in that paper have a worse resolution (by a factor of $\approx 2.0$) and a larger maximum-recoverable-scale (by $\approx 50$\%) that our Band 7 data, and we prefer to analyse a dataset acquired wiith the same interferometer configuration of the \olineh data. We have however checked that the two methods identifying cores in continuum produce results in reasonable agreement. By comparing the cores found in this appendix and in \cite{Sanhueza19} (figure not shown here), 10 out of the 11 cores we identify have correspondence to structures seen in Band 6. In the field-of-view where the two datasets overlap, \cite{Sanhueza19} found more cores, also due to the better sensitivity of their dataset. However, several of the \hhdp-identified cores (5 out of 22) still have no clear correspondence with continuum-identified structures, and our conclusion that continuum and line morfologies are different still holds. }

\section{Results of the spectral fit of the \olineh transition in each core\label{AllCoresMaps}}
 Figs. \ref{AllCores_1} to \ref{AllCores_4} present the maps of the best-fit parameters obtained with \textsc{mcweeds} in each core. Concerning the linewidths, here we show the $FWHM$ maps, which is the actual free parameter used in the fitting procedure. 
 
\begin{figure}[!h]
\centering
\includegraphics[width=.9\textwidth]{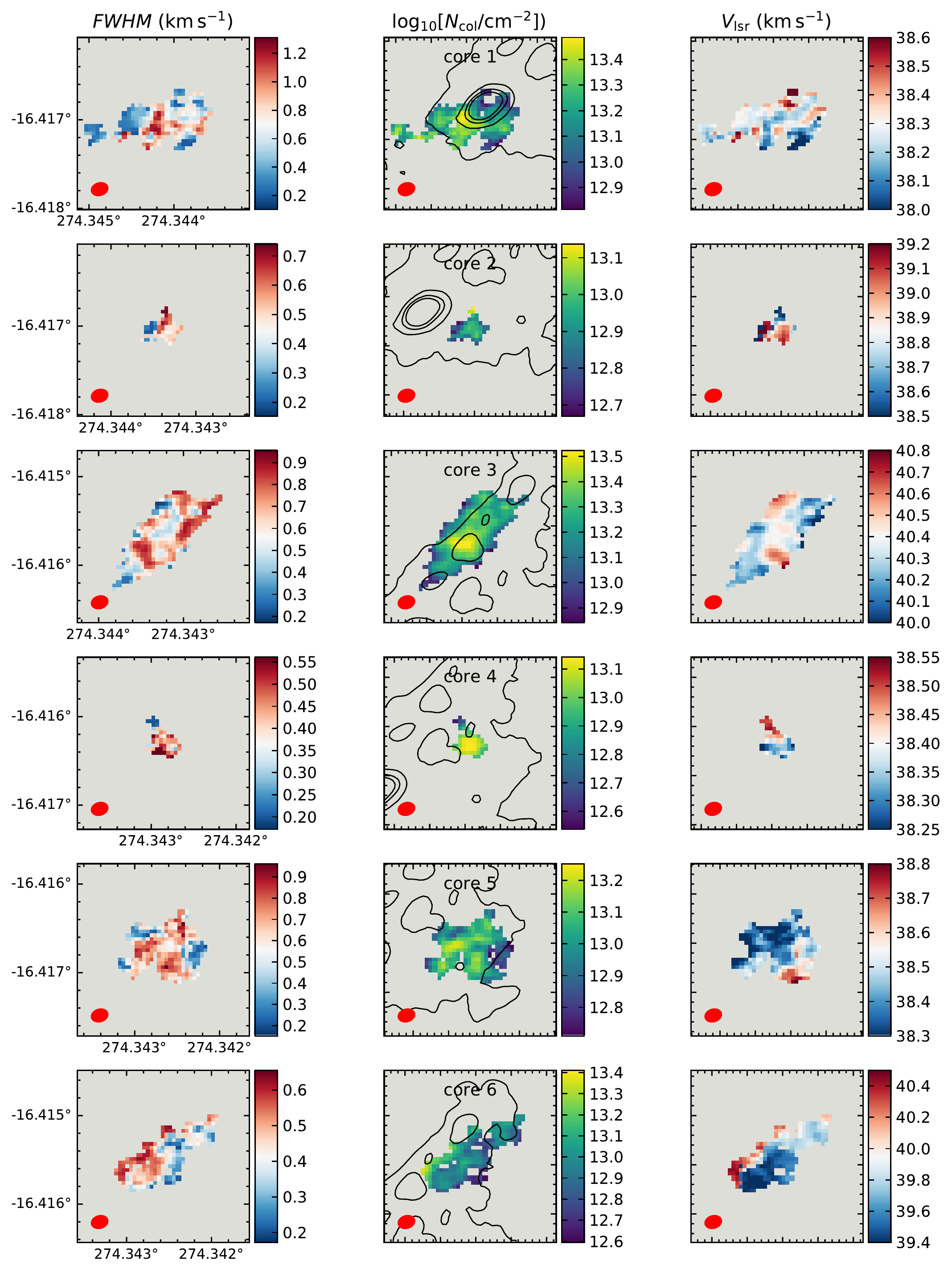}
\caption{Maps of best-fit parameters obtained with \textsc{mcweeds}. The columns are $FWHM$, \ncol, and \vlsr, from left to right. The core label is indicated at the top of each central panel. The contours show the continuum emission at levels from $2$ to $11\sigma$, in steps of $3\sigma$. 
\label{AllCores_1}}
\end{figure}

\begin{figure}
\centering
\includegraphics[width=.9\textwidth]{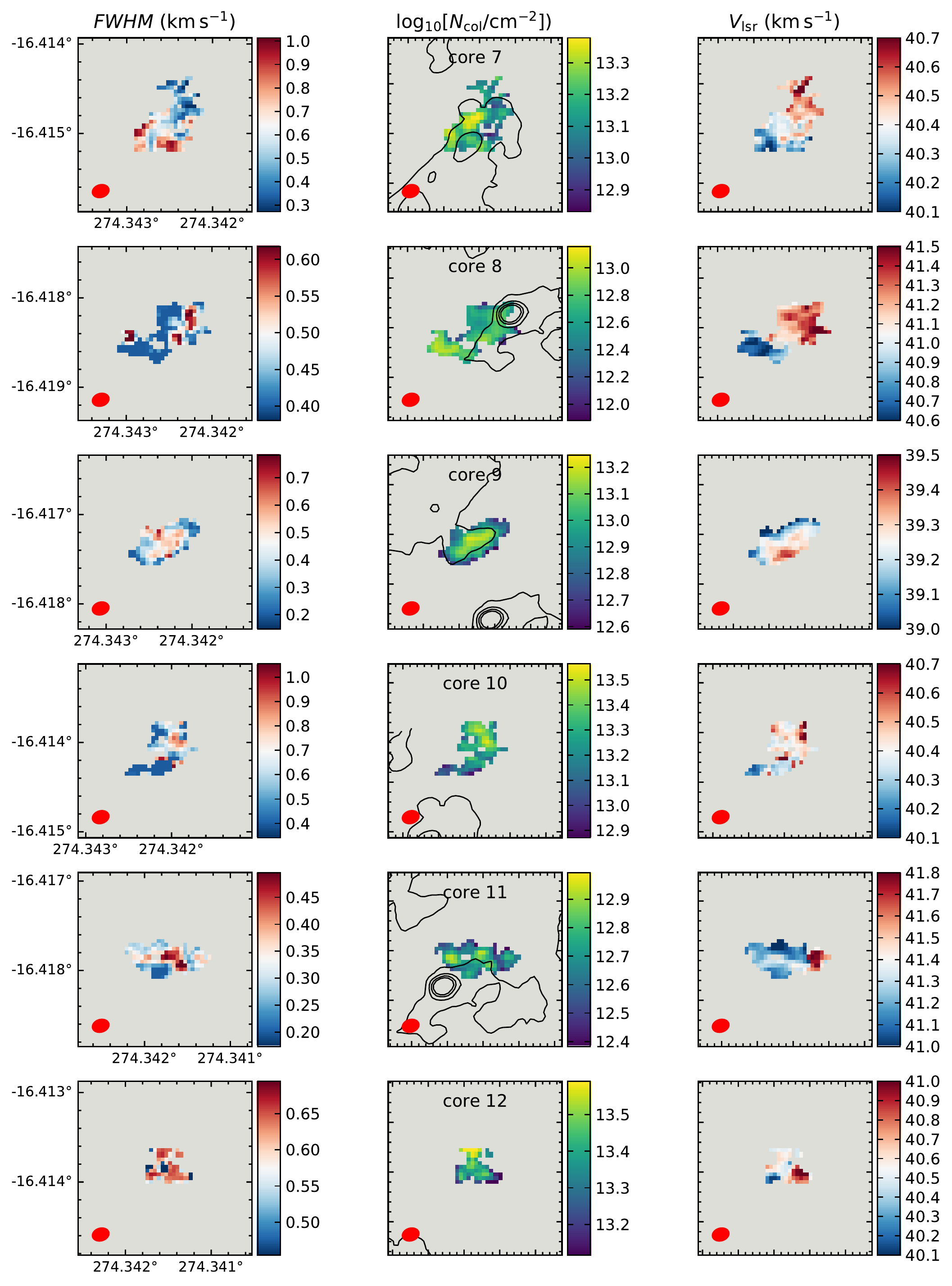}
\caption{Continuation of Fig. \ref{AllCores_1}.
\label{AllCores_2}}
\end{figure}

\begin{figure}
\centering
\includegraphics[width=.9\textwidth]{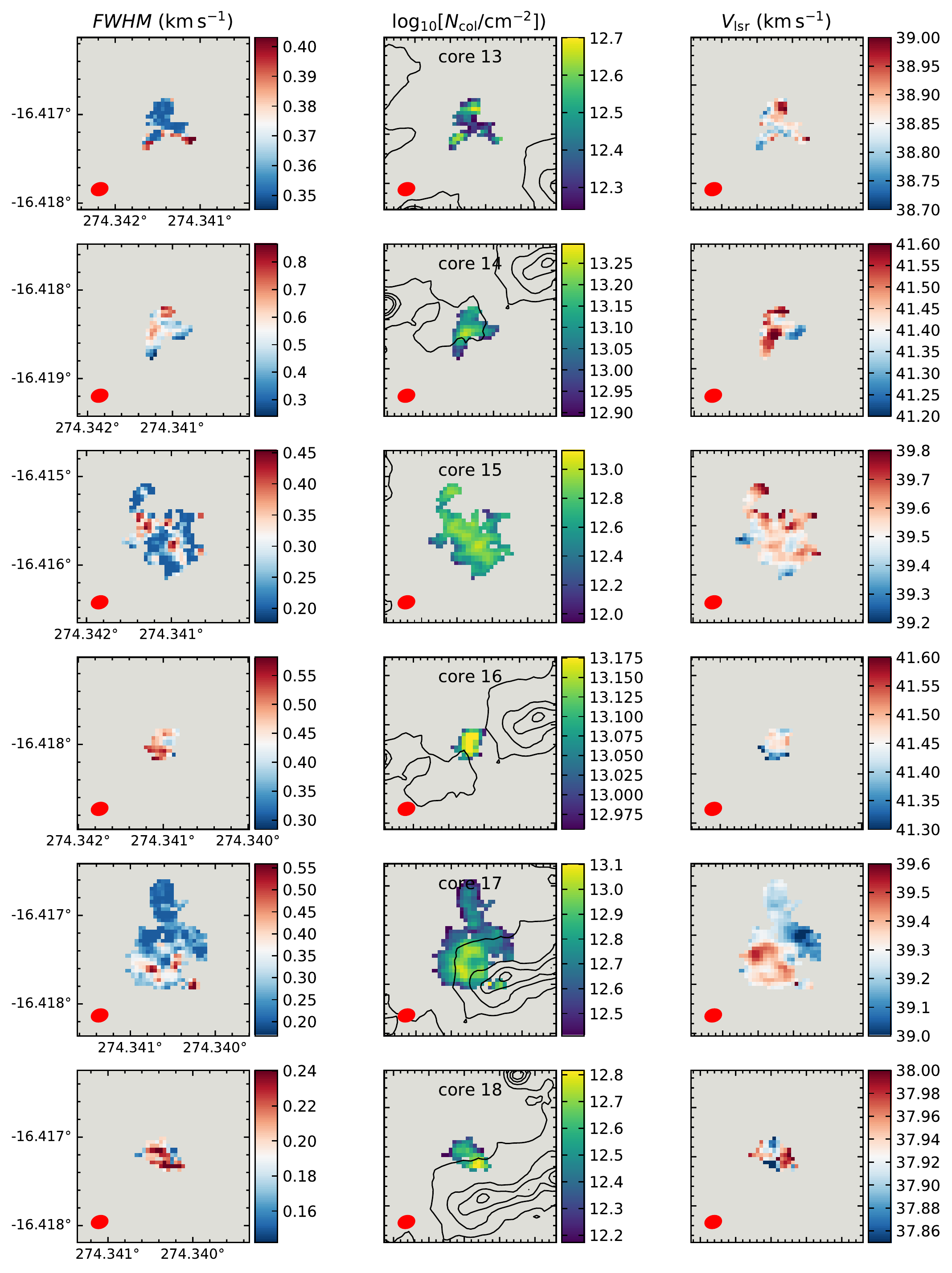}
\caption{Continuation of Fig. \ref{AllCores_1}.
\label{AllCores_3}}
\end{figure}

\begin{figure}
\centering
\includegraphics[width=\textwidth]{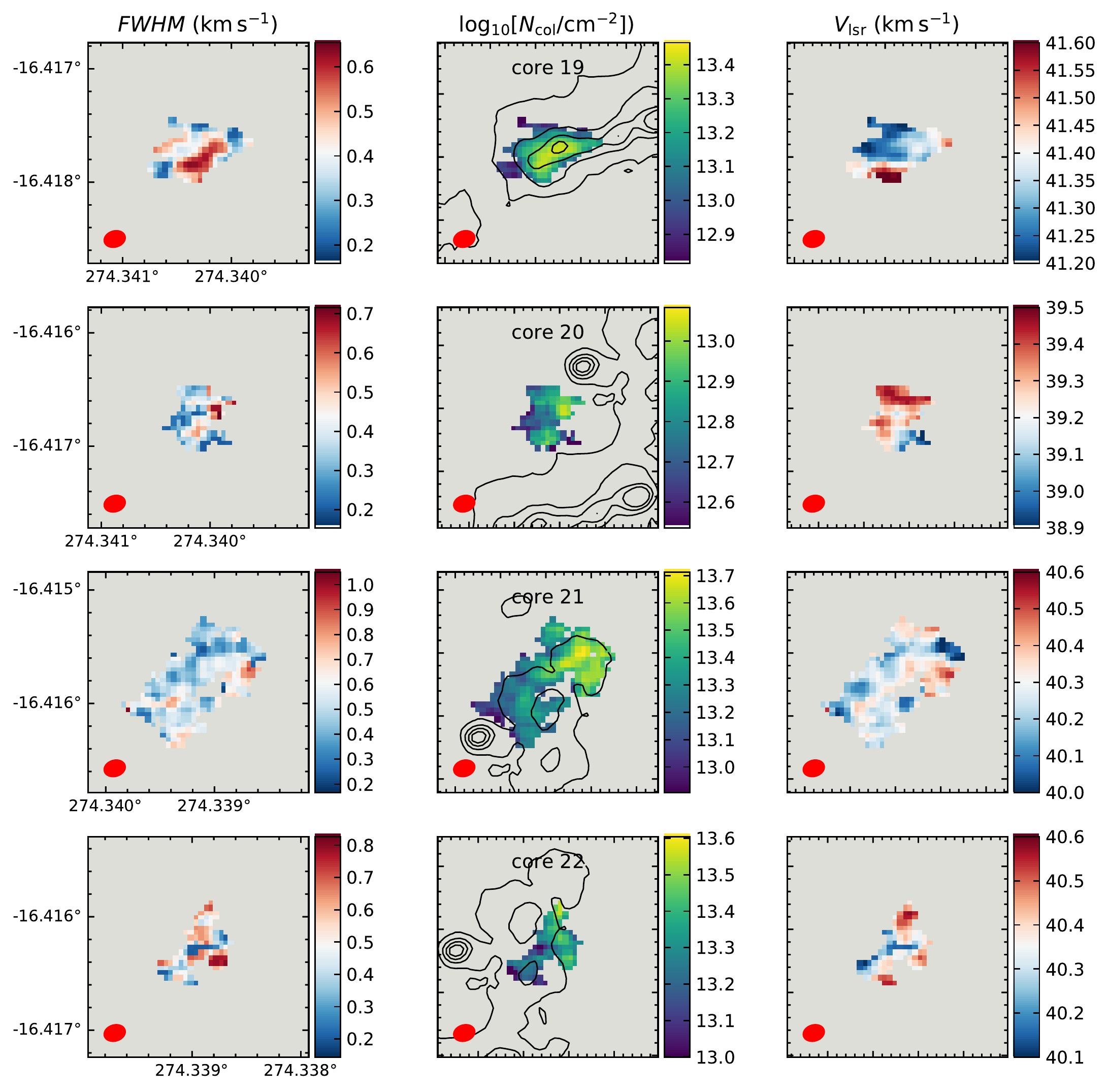}
\caption{Continuation of Fig. \ref{AllCores_1}.
\label{AllCores_4}}
\end{figure}
\clearpage

\section{Opacity and missing flux of the \olineh line \label{missingFlux}}
In order to estimate the opacity of the \olineh line, we make use of the equation: 
\begin{equation}
\tau_\nu = - \ln{1 - \frac{T_\mathrm{b}}{J_\nu(T_\mathrm{ex})- J_\nu(T_\mathrm{bg})}} \; , 
\label{tau}
\end{equation}
where $J_\nu(T)$ is the equivalent Rayleigh-Jeans temperature at the frequency $\nu$ and temperature $T$, and $T_\mathrm{bg}=2.73 \rm \, K$ is the background temperature. In the ALMA data, the brightness temperature peaks at $2\, \rm K$. Using $\text{\tex} = 10 \, \rm K$, Eq. \ref{tau} yields $\tau_\nu \approx 0.8$. Even in the brightest part of the emission, hence, the line is only moderately optically thick. {With this information, we can also estimate by how much the linewidths would be overestimated towards the positions of the source with the highest optical depth. To do so, we make use of Eq. 52 of \cite{Burton92}:
\begin{equation}
\sigma_{\mathrm{obs}} = \frac{\sigma_0} {\sqrt{\ln(2)} } \left \{ \ln  \left [    \frac{\tau_\nu} {       \ln \left ( \frac{2}{1+ e^{-\tau_\nu}} \right) }    \right ]  \right \}^{\frac{1}{2}}  \; ,
\end{equation}
which allows to infer the observed velocity dispersion $\sigma_\mathrm{obs}$ from the intrinsic one $\sigma_0$ given the line opacity $\tau_\nu$. Using the maximum value for the opacity just found, we estimate that in the most optically thick parts of the source the \olineh linewidth is overestimated by 15\% .  }
\par
\begin{figure}[h]
\centering
\includegraphics[width=.6\textwidth]{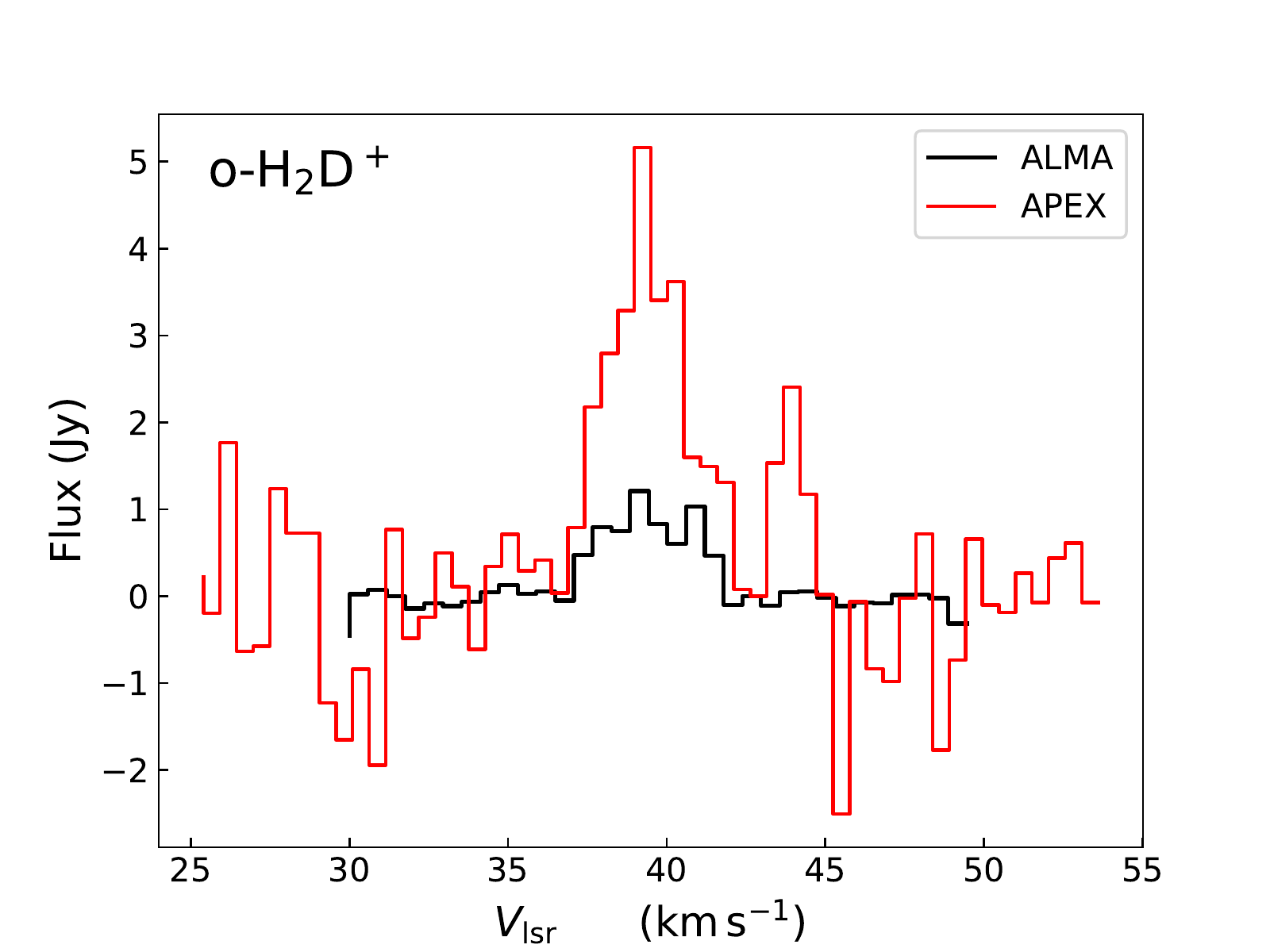}
\caption{Comparison of the \olineh spectra obtained towards AG14 with the APEX single-dish telescope (red histogram) and the ALMA interferometer (black histogram). See the text for the technical details about how the spectra have been extracted. \label{ALMA_APEX}}
\end{figure}

The ALMA Band 7 data lack of Total Power observations, which is crucial to recover the large-scale emission from the source. In order to quantify if and how the observations are affected by filtering-out, we compare the \olineh spectra observed with the APEX single-dish telescope towards AG14 \citep{Sabatini20} with the ALMA data in Fig. \ref{ALMA_APEX}. The APEX data have been converted in flux unit using the gain\footnote{Listed at \url{http://www.apex-telescope.org/telescope/efficiency/}} $G_\mathrm{APEX} = 40 \, \rm Jy \, K^{-1}$. The ALMA data instead have been integrated over an area equal to the beam size of the single-dish ($\theta_\mathrm{APEX} = 16''.8$), and smoothed to the same spectral resolution. \par Figure~\ref{ALMA_APEX} shows that the interferometer is recovering only $\sim$ one fifth of the emission. The missing flux arises from the large scales, since the emission is more extended that the maximum recoverable scale of the telescope in this configuration ($\theta_\mathrm{MRS}\approx 20''$, as was already noted for AG351 and AG354 in \cite{Redaelli21}. However, the ALMA data do not usually present anomalous line shapes. Furthermore, the core identified by \textsc{scimes} are significantly smaller than  $\theta_\mathrm{MRS}\approx 20''$. We hence conclude that the missing flux problem does not affect significantly the analysis of the present work.

\section{\nnhp fitting and full results of \textsc{acorns} clustering \label{app:fit}}
\begin{figure}[!b]
\centering
\includegraphics[width=.8\textwidth]{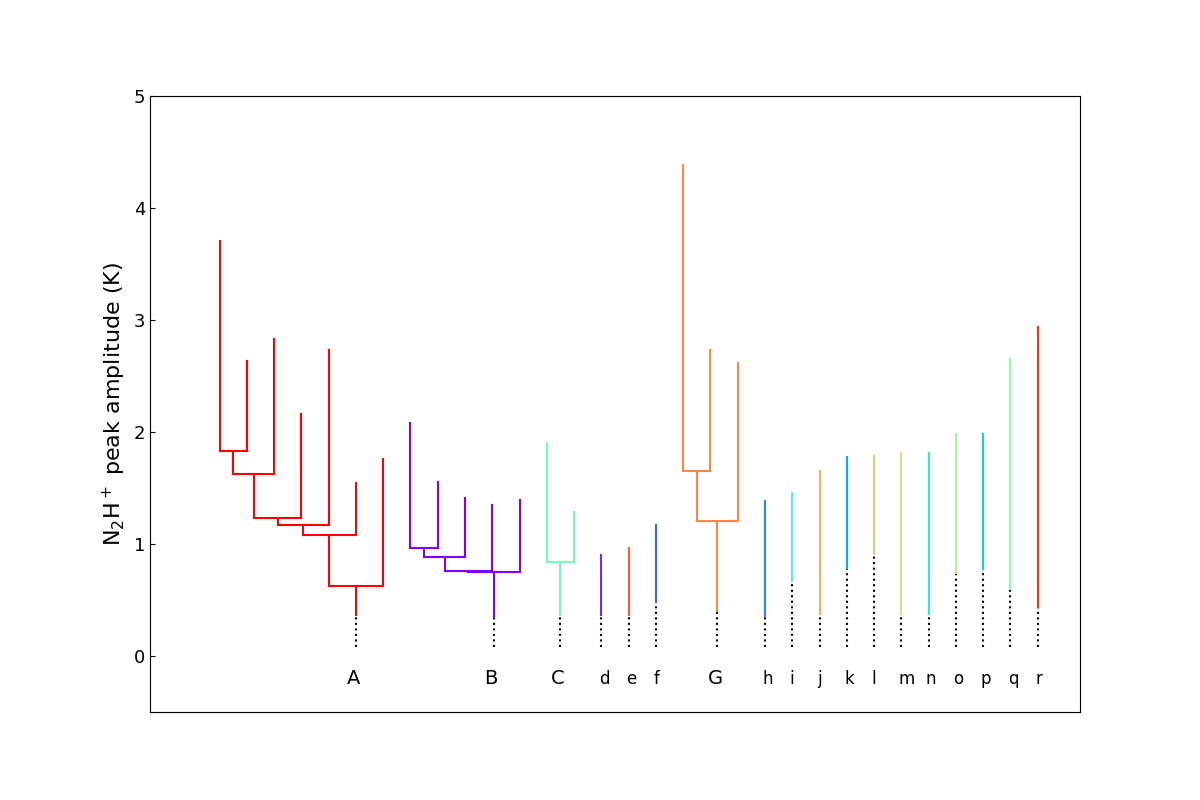}
\caption{Results of the \textsc{acorns} clustering algorithm run on the results of the multi-component Gaussian fitting of the \nnhp (1-0) transition, shown graphically as a dendrogram. The different colours represent distinct trees, labelled with letters, and they correspond to the structures shown in Fig. \ref{acorns_full}. The four trees labelled with capital letters contain more than 70\% of all data, and 80\% of the total flux.  \label{acorns_tot}}
\end{figure}
The multi-component Gaussian fit of the \nnhp isolated hyperfine transition, performed with the \textsc{pyspeckit} package, have nine free parameters in total: \vlsr, \sigmav, and $T_\mathrm{peak}$ values, times three Gaussian components. To improve the code convergence, we first masked pixels with $\rm S/N < 10$ in peak intensity. This choice leaves 5387 positions (55\% of the total) unmasked, which however still cover the whole \ohhdp FoV. We limited the space of the parameters as follows: $V_\mathrm{lsr} \in [36; 43]$\kms; $T_\mathrm{peak} > 0 \,$K; $\sigma_\mathrm{V} \in [0; 2.5]$\kms. Due to the large gradients of the free parameters over the map, the fitting routine does not converge everywhere. After a first procedure, we hence selected the spectra with residuals $> 2 \sigma $ ($1 \sigma = 120 \, \rm mK$; the $rms$ of the residuals is computed in the velocity range $[34.6; 43.5]$\kms), and we performed a second fit, adjusting the initial guesses on the free parameters. We checked the residuals after this second fitting routine, and their $rms$  is found to be $< 3 \sigma$. We further masked pixel-per-pixel velocity components for which the fit did not converge, or with large uncertainties (e.g. $\sigma_{T_\mathrm{peak}}>1 \, $K).
\par
\begin{figure}[!h]
\centering
\includegraphics[width=.9\textwidth]{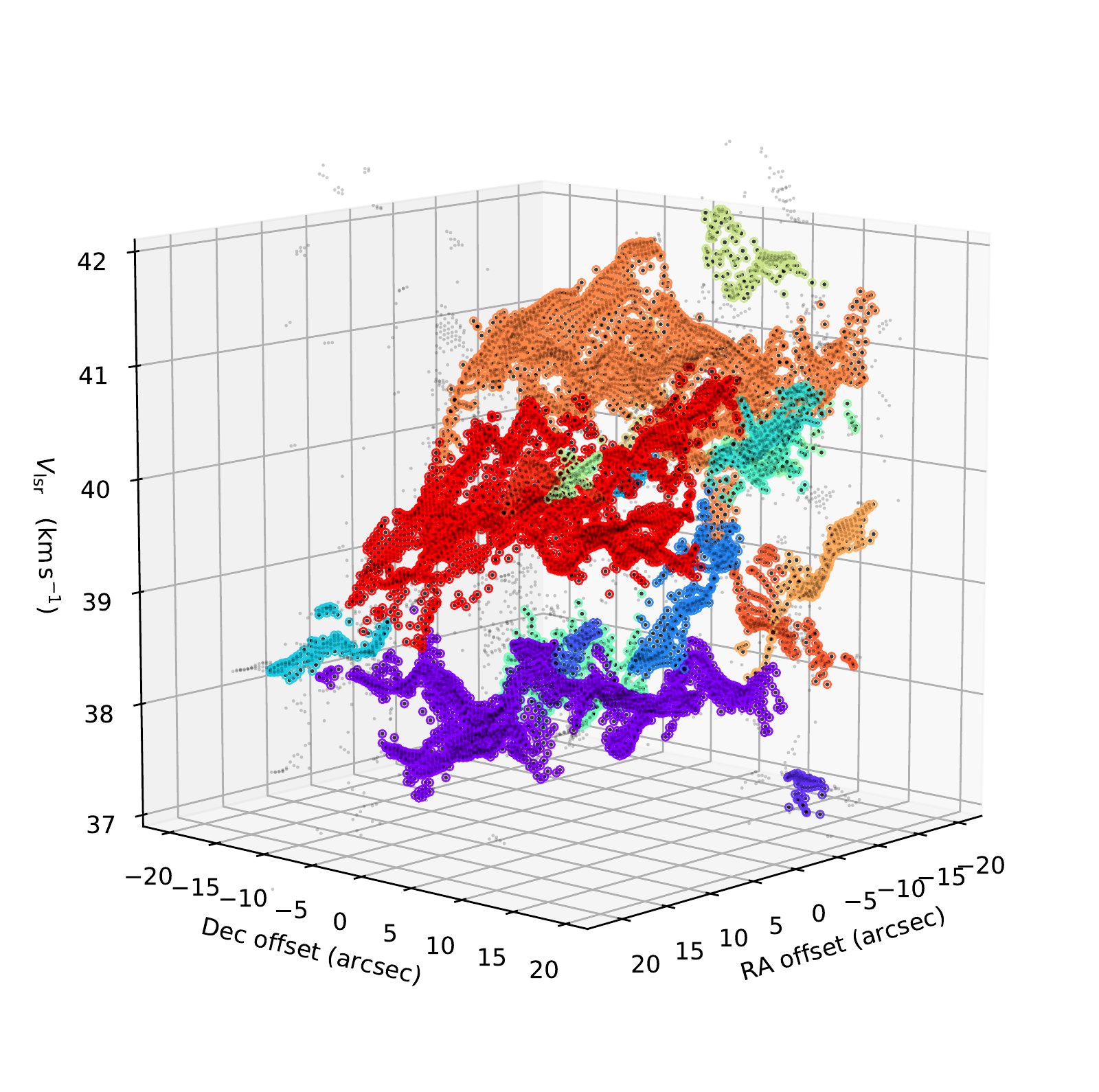}
\caption{Ppv diagram with the complete set of trees identified by \textsc{acorns}. The coordinates axes are expressed as offsets (in arcsec) with respect to the position $\text{RA} =\rm 18^h 17^m 22.0^s $, $\rm Dec = -16\text{\degree} 25' 01''.7$. Each colour represent a different tree, and have correspondence in the dendrogram shown in Fig. \ref{acorns_tot}. Grey points do not belong to any cluster.
\label{acorns_full}}
\end{figure}

The best-fit results of the Gaussian fitting routine are fed to the clustering alorithm \textsc{acorns}. Unlike other similar codes, \textsc{acorns} uses the spectra linewidth as a further parameter to build the cluster hierarchy, and it is overall able to distinguish structures overlapping in ppv space better than other algorithms, which turns helpful for the crowded kinematics of AG14 (see also Appendix B in \citealt{Henshaw19} for further details on the comparison between different algorithms). We select the following clustering criteria in \textsc{acorns}:
\begin{enumerate}
\item Clusters must have a minimum size of 1.5 ALMA beam (to ensure that all the structures found are marginally resolved);
\item They must be separated in velocity less than spectral resolution of the data-cube;
\item The maximum separation in velocity dispersion ($FWHM$) is less than the gas thermal velocity at 10$\,\rm K$ ($0.19\,$\kms);
\item The minimum height of an independent cluster is $3\sigma$, and the stop criteria is set to $5\sigma$ ($1\sigma=0.12\, \rm K$).
\end{enumerate}
After a first run, the code performs a second cycle of clustering, when we relax the criteria by 30\%, which further helps building the hierarchical structure according to the prescriptions of \textsc{acorns}. At the end, the algorithm is able to cluster  87\% of the data-points, and it finds 18 trees, shown in Fig. \ref{acorns_tot} as a dendrogram, {and in Fig. \ref{acorns_full} in ppv space}. The large majority (more than 70\%) of clustered data-points belongs to only four structures, which also contain $\approx 80$\% of the total flux (A, B, C, and G as labelled in Fig. \ref{acorns_tot}). The remaining clusters contain less than 3\% of the data-points each. The analysis of Sect. \ref{kinematics} hence focuses on these four trees.
\par
{We now discuss why we prefer to use distinct softwares to analyse the \nnhp and \ohhdp data. \textsc{scimes} is optimise to work with low-to-medium S/N data, such as the \ohhdp ones. Furthermore, using it ensures a proper comparison with the results of \cite{Redaelli21}, which in turn allows to obtain a larger sample for instance regarding the core masses. \textsc{acorns}, on the other hand, represents a better choice to analyse the \nnhp data, first of all because it has less problems to disentangle crowded spectra such as the ones in AG14.  \textsc{scimes} in fact works on the observed ppv datacubes, and it performs better when the multiple velocity components are well separated in velocity space, as in the \ohhdp data, where these components are separated by $\approx~1\,$\kms and they are narrow ($\sigma_\mathrm{V}=0.3\,$\kms). On the contrary, the \nnhp spectra are more crowded, with more velocity components, and some of these components have significantly broader lines ($\sigma_\mathrm{V}=0.6-1.0\,$\kms). Due to these features, \textsc{scimes} is not able to disentangle them, as demonstrated by a test run of the software that we performed on the \nnhp datacube. \textsc{acorns} instead is able to perform this task because it works on decomposed data. There is also another important difference, in that \textsc{acorns} performs the clustering also in velocity dispersion space. The \olineh linewidths span a much smaller range ($\approx 0.2-0.4\,$\kms) with respect to the \nnhp ones ($\approx 0.3-1.5\,$\kms), and therefore this extra constraint helps even more in disentangling the \nnhp complex kinematics.}



\end{document}